\documentclass[runningheads]{llncs}
\usepackage{float}
\usepackage{graphicx}
\usepackage{listings}
\usepackage{pgfplots}
\usepackage{tikz}

\lstdefinelanguage{Lisp}
{
  alsoletter=-,
  morekeywords={assert, declare-const, BitVec, bvsub, bvslt, bvsgt, check-sat, get-model}
}

\lstdefinelanguage{XML}
{
  morestring=[s][\color{red}]{"}{"},
  morestring=[s]{>}{<},
  morecomment=[s]{<?}{?>},
  stringstyle=\color{black},
  alsoletter=-,
  morekeywords={rule, pattern, message, summary, severity, id, version, function, noreturn, arg, strz, leak-ignore, not-null, not-uninit}
}

\usepgfplotslibrary{units}
\usetikzlibrary{arrows, positioning}
\usetikzlibrary{shapes}

\tikzstyle{CFGNode} = [
    rectangle,
    text centered,
    draw = black,
	minimum width  = 1cm, 
	minimum height = 5mm
]
\pgfplotsset{compat=1.15}

\tikzset{
  treenode/.style = {align=center, inner sep=0pt, text centered,
    font=\sffamily},
  visited/.style = {treenode, circle, black, draw=black, fill=black,
    text width=1.5em, very thick},
  unvisited/.style = {treenode, circle, black, draw=black,
    text width=1.5em, very thick}
}

\begin{document}
\renewcommand{\thelstlisting}{\arabic{lstlisting}}
\title{Scaling Symbolic Execution to \\Large Software Systems}

\author{G\'abor Horv\'ath\inst{1}\orcidID{0000-0002-0834-0996} \and \\
R\'eka Kov\'acs\inst{1}\orcidID{0000-0001-6275-8552} \and \\
Zolt\'an Porkol\'ab\inst{1}\orcidID{0000-0001-6819-0224}}
\authorrunning{G. Horv\'ath et al.}

\institute{Department of Programming Languages and Compilers \\ 
           E\"otv\"os Lor\'and University, Faculty of Informatics \\
           Budapest, Hungary \\
\email{\{xazax,rekanikolett,gsd\}@caesar.elte.hu}}
\maketitle              

\begin{abstract}
Static analysis is the analysis of a program without executing it,
usually carried out by an automated tool.
Symbolic execution is a popular static analysis technique used both
in program verification and in bug detection software. It works
by interpreting the code, introducing a symbol for each value unknown at
compile time (e.g. user-given inputs), and carrying out calculations symbolically. 
The analysis engine strives to explore multiple execution paths simultaneously, 
although checking all paths is an intractable problem, due to the vast number of 
possibilities.

\hspace{0.5cm}We focus on an error finding framework called the Clang Static 
Analyzer, and an infrastructure built around it named CodeChecker.
The emphasis is on achieving end-to-end scalability.
This includes the run time and memory consumption of the analysis,
bug presentation to the users, automatic false positive suppression,
incremental analysis, pattern discovery in the results, and usage in continuous
integration loops. We also outline future directions and open
problems concerning these tools.

\hspace{0.5cm} While a rich literature exists on program verification software, error 
finding tools normally need to settle for survey papers on individual 
techniques. In this paper, we not only discuss individual methods, 
but also how these decisions interact and reinforce each other, creating a system 
that is greater than the sum of its parts.
Although the Clang Static Analyzer can only handle C-family languages, 
the techniques introduced in this paper are mostly language-independent and applicable to 
other similar static analysis tools.
\end{abstract}

\section{Introduction}

Maintenance costs of a software increase with the size of the
codebase. Static analysis has a great impact on 
reducing expenses of complex software~\cite{Zhivich2009}.
For example, compiler warnings rely on various static analysis methods. Moreover,
compilers can detect more and more optimization possibilities
statically and these optimizations make it possible to develop software using
high-level language features. Static analysis, however, is also a great approach
for finding bugs and code smells~\cite{Bessey2010}. The earlier a bug is detected,
the lower the cost of the fix~\cite{fixcost}. This makes static analysis a useful and
cheap supplement to testing, especially as some properties of the code such as 
compliance with coding conventions cannot be tested.

\textit{Testing} is a useful method for catching programming errors but is rarely exhaustive.
Critical corner cases can be easily left uncovered by tests. As famously stated by
Dijkstra: ``Program testing can be used to show the presence of bugs,
but never to show their absence." \textit{Dynamic analysis} is a great
supplement to testing: we run the program with additional runtime checks for errors.
But it has the same shortcomings when it comes to coverage and corner cases. 
One way to cover additional cases is to use \textit{random inputs} \cite{godefroid2012sage}.
However, let's consider the following code snippet (Listing~\ref{lst:sage-testcase}):

\vspace{10pt}
\begin{lstlisting}[caption={An unfavorable test case for random testing.}, captionpos=b, label={lst:sage-testcase}][frame=tlrb]
void f(int x) {
    if (x == 472349) {
        abort();
    }
    // ...
}
\end{lstlisting}
\vspace{5pt}

\noindent While this code is somewhat artificial, it is possible to see that we 
have a very low probability to hit the \texttt{abort()} call during random testing. 

One of the biggest advantages of static analysis is that the analysis can cover cases
developers did not consider. It can provide an elegant solution to systematically
explore interesting execution paths without concrete inputs, using information
inferred from the source code. Moreover, testing (unless exhaustive) cannot prove
the lack of a certain error, while static analysis may be able to prove the
freedom of a class of errors.

Unfortunately, paraphrasing Rice's theorem~\cite{Rice:53} from '53: 
all non-trivial properties of a program are undecidable at compile time. 
Thus, static analysis tools often over-approximate or under-approximate the behavior
of a program. Consequently, such tools may report false errors (called 
\textit{false positives}) or they might miss some real problems 
(called \textit{false negatives}). While verification tools aim to catch all errors at 
the cost of having a large number of false positives, industrial bug-finding tools
aim to have a low false positive rate at the cost of missing some 
true errors. Bug reports need to be reviewed by developers one-by-one
in order to be corrected. If the tool presents an overwhelming amount of false 
warnings to the developer, it becomes cumbersome to use, and developers eventually 
lose their trust and interest in the tool.

This paper is structured as follows. Section~\ref{overview} gives an overview of the most popular
static analysis methods, Section~\ref{symbolicexecution} explores symbolic execution - as the technique we
personally focus on - in depth. Section~\ref{tooling} details the infrastructure built around
the static analysis tools we use. Section~\ref{testing} describes some of the challenges we faced while
testing our analyzers and the solutions we found. Finally, Section~\ref{summary} summarizes
and concludes the paper.

\section{Overview of methods}
\label{overview}

This section will summarize methods that are widely used in static analysis software.
Due to space constraints, we do not intend to give an in-depth overview of these methods nor cover all
available techniques. More specifically, we do not cover formal methods that are 
used for proving program correctness. All of the mentioned approaches are
fully automated, unlike model checking and deductive methods. The user does not
need to supply any invariants nor any description of the semantics of the program, 
everything is derived from the source text.

\subsection{Matching tokens}

Early versions of CppCheck~\cite{marjamaki} are a good example of using regular
expressions to match specific token combinations in the source code. First, the tool 
preprocesses the source file and tokenizes its text, as tokens are easier to process 
in later phases than raw text. The user can write simple rules using regular expressions 
that are matched against the token stream. A sample rule can be seen on Listing~\ref{lst:cpp-check-example}.

In most of the general-purpose programming languages there are many ways to write code 
with the same semantics. CppCheck does many transformations on the token stream to 
simplify the rules written by the user. This process is called \textit{canonicalization}. 
One example is to always use $<$ or $\leq$ operators by eliminating 
$>$ and $\geq$ operators by swapping their operands.

\vspace{10pt}
\begin{lstlisting}[frame=tlrb, language=xml, caption={An example rule to find redundant \texttt{null} checks in CppCheck.}, captionpos=b, label={lst:cpp-check-example}]
<?xml version="1.0"?>
<rule version="1">
  <pattern>if \( p \) { free \( p \) ; }</pattern>
  <message>
    <id>redundantCondition</id>
    <severity>style</severity>
    <summary>It is valid to free a NULL pointer.</summary>
  </message>
</rule>
\end{lstlisting}
\vspace{5pt}

There are several disadvantages of relying on a token stream for writing checks.
First of all, it is hard to reverse-engineer the structure of the code, such as the
precedence of operations, without actual parsing. Second, no type information is available,
making certain checks hard to write. Finally, this representation also misses basic 
information such as which function is being called. In C++, there are complex rules on
how to do overload resolution, but this representation makes it impossible to determine 
the exact callee.

The main advantage of matching tokens is its simplicity. Such checks are easy to implement and very fast to run.

\subsection{Matching the abstract syntax tree}

The \textit{abstract syntax tree} (AST) is a representation of the source code that also
encodes structural information. We not only access tokens but we also know their roles 
and relationships. For example, the comma token in C++ can either be a separator 
for function arguments or the sequence operator. If we only have a token stream, we 
have to deal with such ambiguities, but once we have access to the AST, we know 
exactly what the role of a token is.

Often the AST is decorated with type information. This is important for desugaring
type aliases and resolving calls to overloaded functions. In C++, it is also 
beneficial to have the code instantiated from templates in the AST.

This representation is suitable for catching many common errors, such as space allocation 
on the heap using the wrong size, and it proved to be strong enough to detect some misuses of the 
STL API~\cite{stlmatcher}. Such syntactic checks are usually very efficient and it is 
viable to run them in the editor. Some rules can even be implemented using a single traversal 
of the AST. Clang Tidy~\cite{clangtidy} uses AST matching for most of its checks. 

The disadvantage of AST matching is that we cannot reason about 
possible values of variables or about dependencies of expressions. 

\subsection{Abstract interpretation}

Programs are rather complex structures, thus reasoning about them is hard. 
We may use abstractions to simplify the analyzed program and check for certain 
properties separately. For example, instead of trying to reason about the exact 
values a variable can hold, we can reason about the signedness of those values. 
Similarly, instead of trying to enumerate feasible execution paths of a program 
(which is undecidable), we can assume that all of the paths are feasible.

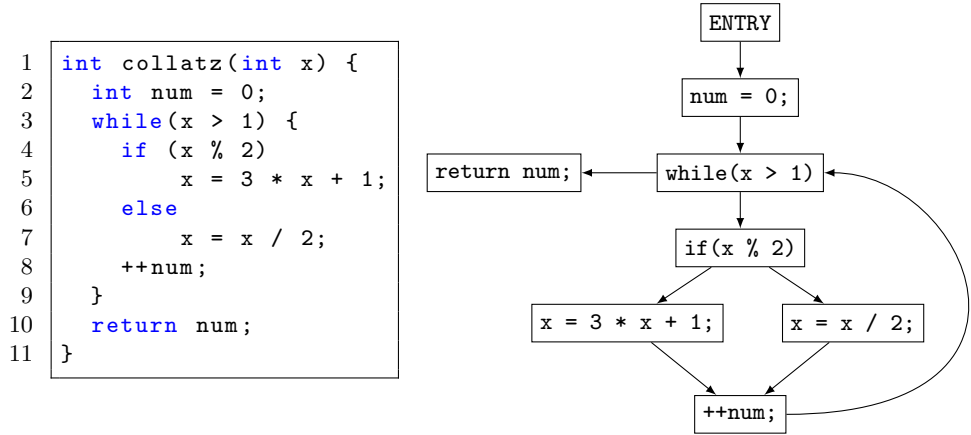
\begin{figure*}
\noindent\begin{minipage}{.36\textwidth}
\begin{lstlisting}[frame=tlrb]{Name}
int collatz(int x) {
  int num = 0;
  while(x > 1) {
    if (x % 2)
        x = 3 * x + 1;
    else
        x = x / 2;
    ++num;
  }
  return num;
}
\end{lstlisting}
\end{minipage}\hfill
\begin{minipage}{.6\textwidth}
\begin{tikzpicture}[
	level distance = 1cm,
	level 4/.style = {sibling distance=3cm},
	edge from parent/.style = {
		draw,
		-latex
	}
]
\node [CFGNode]{\texttt{ENTRY}}
child {node [CFGNode]{\texttt{num = 0;}}
	child {node (whilenode) [CFGNode]{\texttt{while(x > 1)}}
		child {node (ifnode) [CFGNode]{\texttt{if(x \% 2)}}
			child {node (truenode) [CFGNode]{\texttt{x = 3 * x + 1;}}
			}
			child {node (falsenode) [CFGNode]{\texttt{x = x / 2;}}
			}
		}
	}
};
\node (plusplusnumnode) [CFGNode, below = 1.7cm of ifnode] {\texttt{++num;}};
\path[every node/.style={font=\sffamily\small}]
(truenode) edge[-latex] (plusplusnumnode);
\path[every node/.style={font=\sffamily\small}]
(falsenode) edge[-latex] (plusplusnumnode);
\path[every node/.style={font=\sffamily\small}]
(plusplusnumnode) edge[in=0, out=0, in looseness=1.5, out looseness=3
, -latex] (whilenode);
\node (returnnode) [CFGNode, left = 1cm of whilenode] {\texttt{return num;}};
\path[every node/.style={font=\sffamily\small}]
(whilenode) edge[-latex] (returnnode);
\end{tikzpicture}
\end{minipage}
\caption{A C function and its simplified control flow graph.}
\label{fig:CFG}
\end{figure*}

Using such methods we over-approximate the behavior of the program and simulate each
possible execution without using too many resources during the analysis. Typically,
for such analyses we use the \textit{control flow graph} (CFG) of the program. 
The nodes of this graph are \textit{basic blocks}, which are sequences of instructions 
always executed sequentially, while edges are the possible jumps between basic blocks. 
An example CFG can be seen in Figure~\ref{fig:CFG}.

During \textit{abstract interpretation}~\cite{abstractinterpret} we reason about 
the possible values of variables at a certain program point. We can think of a program 
as a sequence of program states and transitions. Representing all the possible program 
states is an intractable problem, thus we use an abstract program state that captures 
some aspects of the variables. The way we represent this abstract program state is 
determined by the \emph{abstract domain}. Some examples of abstract domains in
increasing expressiveness are signs, 
intervals, octagons~\cite{mine2006octagon}, and polyhedra~\cite{singh2017fast}.
For example, if we choose our abstract domain to 
be intervals, a program state will consist of a program point and a mapping from each 
variable to an interval that over-approximates the possible values of the variable at 
the program point described by the state.

For an abstract interpretation algorithm we also need to define how each basic block
transforms abstract values. For example, if we do sign analysis, we need to know
which variables' signs are changed by a basic block. The functions describing this
transformation are called \textit{transfer functions}.

\subsubsection{Flow-sensitive analysis} 
Most of the times we are able to produce the 
values in the domain of the analysis for each basic block. We have a set of initial 
conditions for each basic block and propagate values between basic blocks using 
fixed-point iteration. The reason why such analysis will converge is that
we chose the abstract domain to be a lattice\footnote{A \emph{lattice} is a partially 
ordered set, in which each pair of elements has a unique supremum and a unique 
infinum.}~\cite{birkhoff1940lattice}, and each operation during the analysis
can only decrease the abstract value of a variable. Thus, in the worst case, we will
reach a fixed point where each variable is mapped to the smallest element of the lattice
(often called bottom).

\subsubsection{Path-sensitive analysis} 
The flow-sensitive analysis assumes that every
possible walk over the CFG is a feasible path. That is almost never the case, since some
of the paths can never be taken during execution. Let us consider the following
example (Listing~\ref{lst:path-sensitive}):

\vfill
\pagebreak

\vspace{10pt}
\begin{lstlisting}[frame=tlrb, caption={A function that demonstrates the difference in precision between flow-sensitive and path-sensitive analyses.}, captionpos=b, label={lst:path-sensitive}]
int f(int x) {
  int c = 0;
  if (x > 2) {
    c = 3;
  }
  // ... (c is unchanged)
  if (x > 5) {
    return 9/c;  // Division by zero?
  }
}
\end{lstlisting}
\vspace{5pt}

\noindent A flow-sensitive analysis would report a division by zero error in 
the code above, even though such an error could never occur during runtime, since
each time the value of \texttt{x} is greater than five, the value of \texttt{c} will 
not be zero. One possible improvement is to record \textit{path constraints} for each 
checked execution path. A path-sensitive analysis will use the path constraint and 
SMT solvers to prune (some of) the infeasible paths. In the code above such analysis 
could derive that we never take the \texttt{x > 5} branch when the value of 
\texttt{c} is zero. This method can help reduce the number of false positives 
significantly, but it can also result in a combinatorial explosion of states.

\subsection{Symbolic Execution}

Symbolic execution~\cite{Hampapuram2005,symex,survey} is a path-sensitive static analysis method. 
In abstract interpretation, we over-approximate the behavior of the analyzed 
program and reason about every possible execution. On the other hand, during symbolic 
execution, we only reason about a set of paths, but more precisely. Thus, symbolic 
execution is both an over-approximation (there is a loss of information when we 
represent symbolic states) and under-approximation (we do not cover all of the 
possible execution paths).

\section{Symbolic Execution}
\label{symbolicexecution}

This chapter introduces symbolic execution in depth, concentrating on
methods used by the Clang Static Analyzer~\cite{analyzer} (henceforth referred to as 
\textit{the analyzer}). There are other tools available that use
similar techniques, e.g. Infer~\cite{infer}, which is open-source and free to use.
KLEE~\cite{cadar2008klee} is also open source but it works on a lower
level representation of the code and one of its
purpose is to generate test cases automatically.
A list of proprietary tools include CodeSonar~\cite{codesonar}, 
Klocwork~\cite{klocwork}, and Coverity~\cite{coverity}.

Symbolic execution interprets the source code, assigning a symbol to represent
each unknown value. Calculations are carried out symbolically. During the 
interpretation process, the analyzer attempts to enumerate all possible execution paths.
To represent the internal state of the analysis, the analyzer uses a data structure 
called the \emph{exploded graph}~\cite{Reps95}.
Each vertex of this graph is a (\textit{symbolic state}, \textit{program point}) pair.
A \emph{symbolic state} corresponds to a set of real program states, while the 
\emph{program point} determines the current location in the program, similarly to an 
instruction pointer. The edges of the graph are transitions between vertices.
Memory is represented using a hierarchy of memory regions~\cite{Xu2010}.
The analyzer builds the graph on demand during the analysis using a 
\textit{worklist algorithm} that implements a path-sensitive walk over the CFG.

\noindent The symbolic state consists of 3 components:
\begin{itemize}
\item[\textbullet] \textbf{Environment}: A mapping from source code expressions to symbolic expressions.
\item[\textbullet] \textbf{Store}: A mapping from memory locations to symbolic expressions.
\item[\textbullet] \textbf{Generic data map}: A data structure where the analysis engine and checks store domain-specific information.
\end{itemize}

During the execution of a path, the analyzer collects constraints on
symbolic expressions called \textit{path constraints}. There is a built-in 
constraint solver that represents these constraints using a disjunction of 
ranges. This solver can reason about integers and pointers in linear expressions. 
It is also possible to use the \emph{Z3 Theorem Prover}~\cite{deMoura2008} as an 
external solver. The groundwork of adding support for more solvers is already being 
laid down, with the planned ability to use multiple ones at once for the same analysis.
As mentioned earlier, these constraints are used to skip the analysis of 
infeasible paths. The constraint solver, however, can also be
utilized by the checks to query certain information about the program states.
This functionality can be used, for instance, to detect array out-of-bounds
accesses or division by zero errors.
\vspace{-15pt}

\begin{figure}[H]
\noindent\begin{minipage}{.28\textwidth}
\begin{lstlisting}[frame=tlrb]{Name}
void g(int b,
       int &x) {
  if (b)
    x = b+1;
  else
    x = 42;
}
\end{lstlisting}
\end{minipage}\hfill
\begin{minipage}{.65\textwidth}
\begin{tikzpicture}[scale=0.8, transform shape]
\draw  (-2,2.5) rectangle (1,1) node[pos=.5, rectangle split,rectangle split parts=3] {b: $\$b$, x: $\$x$\nodepart{second} \$b : [IMIN, IMAX]\nodepart{third} \$x : [IMIN, IMAX]};
\draw  (-4.5,0) rectangle (-1.5,-1.5) node[pos=.5, rectangle split,rectangle split parts=3] {b: $\$b$, x: $\$x$\nodepart{second} \$b : [0, 0]\nodepart{third} \$x : [IMIN, IMAX]};
\draw  (-4.5,-2.5) rectangle (-1.5,-3.5) node[pos=.5, rectangle split,rectangle split parts=2] {b: $\$b$, x: $42$\nodepart{second} \$b : [0, 0]};
\draw  (0.5,0) rectangle (5,-1.5) node[pos=.5, rectangle split,rectangle split parts=3] {b: $\$b$, x: $\$x$\nodepart{second} \$b : [IMIN, -1]$\ \cup\ $[1, IMAX]\nodepart{third} \$x : [IMIN, IMAX]};
\draw  (0.5,-2.5) rectangle (5,-3.5) node[pos=.5, rectangle split,rectangle split parts=2] {b: $\$b$, x: $\$b+1$\nodepart{second} \$b : [IMIN, -1]$\ \cup\ $[1, IMAX]};
\draw [->](-0.5,1) node (v1) {} -- (-3,0);
\draw [->](v1) -- (2.5,0);
\draw [->](-3,-1.5) -- (-3,-2.5);
\draw [->](2.5,-1.5) -- (2.5,-2.5);
\end{tikzpicture}
\end{minipage}
\caption{A simplified version of the exploded graph built during the symbolic execution of a simple function. The right-hand side of the graph represents the \textit{true} branch of the \texttt{if} statement, while the left-hand side describes the \textit{false} branch.}
\label{fig:exploded}
\end{figure}
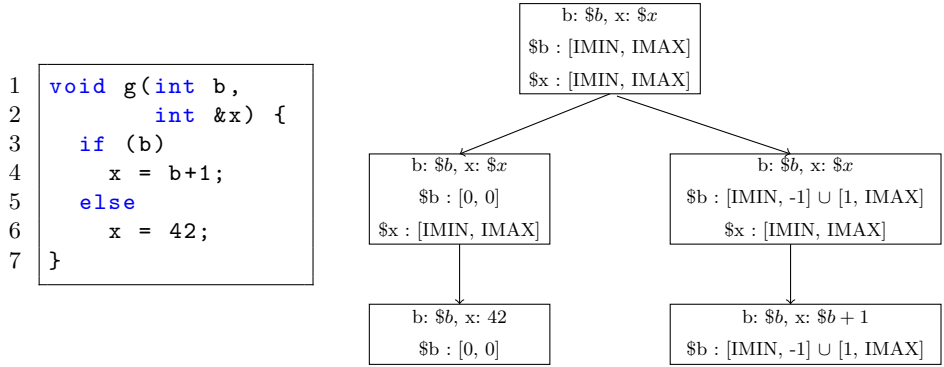

\vfill
\pagebreak

An example analysis can be seen along with its simplified exploded graph on 
Figure~\ref{fig:exploded}. Each box in the exploded graph represents a symbolic 
program state. The first line of the box is the \textit{store}, which holds mappings 
between memory regions and symbolic expressions. The other lines
represent path constraints over symbols collected during exploration. 
We omitted any remaining components for brevity.

In function \texttt{g}, the values of \texttt{b} and \texttt{x} are initially unknown. 
These values are represented in the analysis by symbols \texttt{\$b} 
and \texttt{\$x}, which can take any arbitrary value that their type permits.
Since \texttt{b} acts as the condition of the \texttt{if} statement,
its range of values can be divided into two interesting subsets: \texttt{\$b} $\in$ [0, 0],
which leads to the \texttt{then} branch, and \texttt{\$b} $\in$ [IMIN, -1] $\cup$ [1, IMAX],
which leads to the \texttt{if} branch. \texttt{IMIN} and \texttt{IMAX} represent the
minimal and maximal integer value available on a given platform, respectively.
As the analysis continues, on the execution path where the value of \texttt{b} is 
assumed to be zero, we later discover that the value of \texttt{x} is the constant 
\texttt{42}. The symbol \texttt{\$x} is no longer needed on this path. 
On the other path, the value of \texttt{b} can be anything
but zero. Later we discover that the value of \texttt{x} is one larger
than the original value of \texttt{b}. The symbol \texttt{\$x} is no longer needed
on any of the paths, it can be garbage collected.

The real exploded graph generated by Clang for this code snippet consists of 19 nodes 
and each node contains more information than our simplified nodes. A fraction of the 
real graph can be examined in Figure \ref{fig:realexploded} in Appendix A.

Notice that each node in the exploded graph is very similar to its predecessor.
Each transition will only modify a small part of the symbolic program state. Storing
all states in memory separately would cost a lot of memory. The analyzer uses an 
\textit{immutable/persistent AVL tree}~\cite{persistent} to mitigate this 
problem. Using immutable data structures, each time we create a new state, it will 
refer back to its predecessor instead of copying the old one. Thus each element of
the state will be stored only once regardless of how many states are referring to
that element.

Symbolic execution might find the same error on multiple paths. For example, we might
allocate a chunk of memory that we never release, and the function might have multiple
return paths. The analyzer will find the memory leak on every path. Instead of flooding 
the user with all of the paths, the analyzer will only present the shortest one, as seen
on Listing~\ref{lst:deduplication}.

\vspace{10pt}
\begin{lstlisting}[frame=tlrb, caption={The memory leak stemming for never deallocating
memory pointed to by \texttt{p} can be realized on both a shorter and a longer path
through \texttt{f}. The analyzer will only warn on the shortest path to avoid spurious
error reports.}, captionpos=b, label={lst:deduplication}]
int f(int x) {
    int *p = new int;
    if (x > 0)
        return 1; // Warning: memory leak.
    // ...
    return 0;
}
\end{lstlisting}

\vfill
\pagebreak

\subsection{Memory model}

How should the analyzer represent memory? Let us explore this problem through the 
following example (Listing~\ref{lst:memory}):

\vspace{10pt}
\begin{lstlisting}[frame=tlrb, caption={Motivating example for the hierarchical memory model.}, captionpos=b, label={lst:memory}]
struct X { int a, b; };
void unknown(int *);
void unknown2(X *);
void g() {
  X x{0, 2};           // A
  unknown(&x.a);       // B
  int val = 5 / x.a;   // Should we warn?
  unknown2(&x);        // C
  int val2 = 5 / x.b;  // Should we warn?
}
\end{lstlisting}
\vspace{5pt}

\noindent In function \texttt{g}, at program point \texttt{A}, we know the value of 
each local variable including the fields of \texttt{x}. What can we say about 
program point \texttt{B}? We do not know the body of function \texttt{unknown} and 
it is free to modify the value of \texttt{x.a}, since we have passed its address. 
Thus, a division by zero warning is 
likely to be a false positive. In order to avoid such spurious warnings we sometimes 
need to erase some information from the program state. We call this process 
\textit{invalidation}. After the call to \texttt{unknown} the value of \texttt{x.a} 
is no longer known, the analyzer will not warn at the division.
What happens at program point \texttt{C}? Since the \texttt{unknown2} function is free to
modify any of the fields of \texttt{x} we need to invalidate both of them to avoid the
false positive. Thus, we need to model the relation between the memory of the object 
\texttt{x} and its fields \texttt{x.a} and \texttt{x.b}.

\vspace{-10pt}
\begin{figure}
\noindent\begin{minipage}{.2\textwidth}
\begin{lstlisting}[frame=tlrb]{Name}
   a[0]
   a[1].f
\end{lstlisting}
\end{minipage}\hfill
\begin{minipage}{.65\textwidth}
\begin{tikzpicture}[scale=0.8, transform shape]
\draw  (-2,2.5) rectangle (1,2) node[pos=.5, rectangle split,rectangle split parts=1] {Stack Region};
\draw  (-2,1.5) rectangle (1,1) node[pos=.5, rectangle split,rectangle split parts=1] {\texttt{a}: Array Region};
\draw  (-5,0.5) rectangle (-1.5,0) node[pos=.5, rectangle split,rectangle split parts=1] {\texttt{a[0]}: Element Region};
\draw  (0.5,0.5) rectangle (4,0) node[pos=.5, rectangle split,rectangle split parts=1] {\texttt{a[1]}: Element Region};
\draw  (0.5,-0.5) rectangle (4,-1) node[pos=.5, rectangle split,rectangle split parts=1] {\texttt{a[1].f}: Field Region};
\draw [->](-0.5, 2)  -- (-0.5,1.5);
\draw [->](-0.5, 1)  -- (-3,0.5);
\draw [->](-0.5, 1)  -- (2,0.5);
\draw [->](2,0)  -- (2,-0.5);
\end{tikzpicture}
\end{minipage}
\caption{\textit{Left:} An expression representing field \texttt{f} of an object 
that is stored as the second element of array \texttt{a} on the stack. \textit{Right:} 
The region hierarchy describing the memory layout of the expression on the left.}
\label{fig:memregion}
\end{figure}
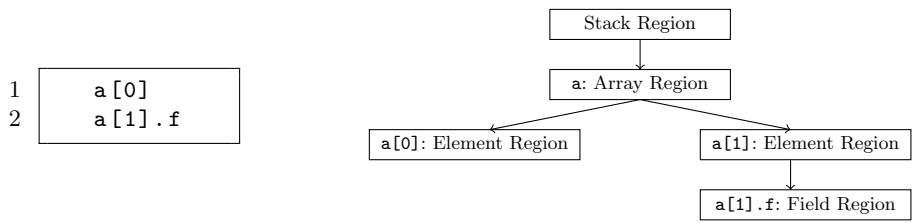
\vspace{-10pt}

The analyzer is using a hierarchical representation for memory regions. 
Figure~\ref{fig:memregion} contains a small snippet of code and its representation 
in the analyzer. We call the top region the \textit{memory space region}. 
It is often useful to know whether a certain variable is allocated on the stack, on 
the heap, or is a global variable. We call the immediate children of the memory 
space regions \textit{base regions}. As an optimization some internal algorithms
in the analyzer are working on base regions. This way these algorithms will not 
need to process the whole sub-tree but only one region for each of the 
corresponding sub-trees.

Note that this representation could be extremely memory-heavy. Imagine an array of  
one million elements where each element is a \texttt{struct} of one hundred fields. 
Creating a tree for one hundred million memory regions would not only consume a lot 
of memory but also take a lot of time. Because of this, the analyzer will not create 
the memory regions upfront. It will only create one region for the array and construct 
child regions lazily each time an element is written to. Until then all of the reads 
to any of the elements will return a default value represented by a default binding. 
For an uninitialized array, this default binding would represent an undefined value.

\subsection{Context sensitivity}

Some errors span across multiple functions. For example, we might mistakenly delete a chunk of
allocated memory once in one function and again in another one. Thus, it is useful to
be able to analyze across function call boundaries. If we analyze a function
only once for all the call sites, we call the analysis \textit{context-insensitive}. 

In \textit{context-sensitive analysis} we differentiate between call sites and 
analyze the function body for each calling context individually. In the Clang Static 
Analyzer we use \textit{inline substitution / function cloning} to achieve 
context-sensitive analysis, meaning that the analyzer will act as if the function 
body was copied to the call site. 

Note that we cannot inline all the functions due to cycles in the function call graph. 
In case the analyzer cannot inline a call, it will handle the function as 
\textit{unknown} and do the necessary invalidations to avoid spurious warnings. 
We call this \textit{conservative evaluation} of a function call.
Also, we do not always know the calling context of a function. There might be functions
without call sites that are entry points to our program, or the call sites are not known
due to separate compilation. Those functions are analyzed without calling contexts and
we call them \textit{top-level functions}. 

In order to improve the false positive ratio, the analyzer will not re-analyze a function
as top-level if it was already analyzed via inlining. The reason is that we always have
more information after inlining from the calling context. We also might end up losing
coverage due to some of the paths not being exercised by the call sites observed by
the analyzer. For example, for the code in Figure~\ref{fig:context}, the analyzer will 
only check \texttt{f} in the calling context of \texttt{g}, thus it will not find the 
division by zero error.

\vfill
\pagebreak

\begin{figure}[h!]
\centering
\begin{minipage}{.42\textwidth}
\begin{lstlisting}[frame=tlrb]
int f(int x) {
    if (x > 0)
        return x;
    return x/0;
}
\end{lstlisting}
\end{minipage}\hfill
\begin{minipage}{.42\textwidth}
\begin{lstlisting}[frame=tlrb]
int g() {
    return f(3);
}
\end{lstlisting}
\end{minipage}
\caption{In the Clang Static Analyzer, an inlined function will not be re-analyzed 
as a top-level function. The analyzer only checks function \texttt{f} in the calling 
context of function \texttt{g}, thus it does not find the division by zero error in 
\texttt{f}.}
\label{fig:context}
\end{figure}
\vspace{-10pt}

\noindent While not re-analyzing certain functions as top-level will introduce false 
negatives, it is also an optimization and reduces the number of false positives.

What functions should be picked for top-level analysis? We can use topological sorting
to pick good candidates. While we cannot do a topological sort on cyclic graphs, we can
use heuristics to have an order with a close-to-minimal number of back edges in
the determined order of vertices. In the future we also plan to experiment with
annotations, so the user can specify which functions should be treated as entry points.

\subsection{Exploration strategy}

The biggest weakness of symbolic execution is its time complexity. The number of explored
paths can be exponential in the branching factor. Moreover, if the analyzed software has
loops, the number of paths might be infinite. Thus, it is not possible to do an exhaustive
search over all paths. Symbolic analysis engines usually have some budgets that decide
when to stop exploring a certain path and when to stop analyzing a top-level function. 
Once we have a budget, we can keep the run time of the analysis within reasonable limits. 

It is not trivial, however, to decide which paths to examine during this limited time. 
Since we use worklist algorithm, we are free to choose which basic block to process next. 
This choice will determine the exploration strategy. The analyzer implements multiple ones: 
\textit{breadth-first search} (BFS), \textit{depth-first search} (DFS), and
\textit{unexplored first search}. 

The default strategy will prioritize basic blocks that have been visited the least 
times so far during the analysis. This helps us maximize analysis
coverage. As we will see later, we also prefer to find bugs on short execution paths as
they are more likely to be true positive, and easier to interpret by 
developers~\cite{Hallem2002}. Thus, we usually favor heuristics that prefer to explore 
multiple short paths rather than a few long ones.

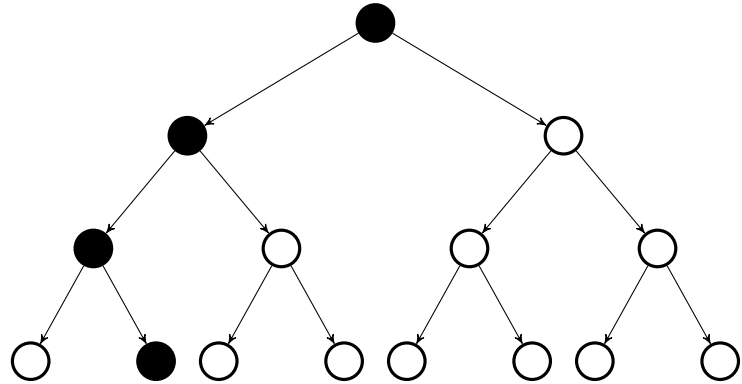
\begin{figure}[!h]
\centering
\begin{tikzpicture}[->,>=stealth',level/.style={sibling distance = 5cm/#1,
  level distance = 1.5cm}] 
\node [visited] {}
    child{ node [visited] {} 
            child{ node [visited] {} 
            	child{ node [unvisited] {} } 
				child{ node [visited] {}}
            }
            child{ node [unvisited] {}
					child{ node [unvisited] {}}
					child{ node [unvisited] {}}
            }                            
    }
    child{ node [unvisited] {}
            child{ node [unvisited] {} 
							child{ node [unvisited] {}}
							child{ node [unvisited] {}}
            }
            child{ node [unvisited] {}
							child{ node [unvisited] {}}
							child{ node [unvisited] {}}
            }
		}
; 
\end{tikzpicture}
\caption{An illustration of the depth-first search (DFS) exploration strategy.}
\label{fig:DFS}
\end{figure}

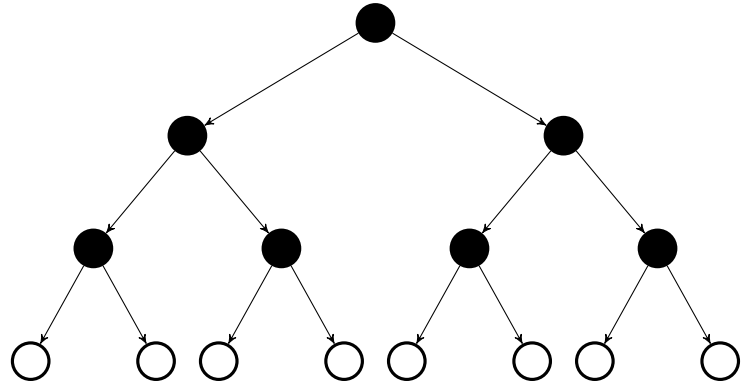
\begin{figure}[!h]
\centering
\begin{tikzpicture}[->,>=stealth',level/.style={sibling distance = 5cm/#1,
  level distance = 1.5cm}] 
\node [visited] {}
    child{ node [visited] {} 
            child{ node [visited] {} 
            	child{ node [unvisited] {} } 
				child{ node [unvisited] {}}
            }
            child{ node [visited] {}
				child{ node [unvisited] {}}
				child{ node [unvisited] {}}
            }                            
    }
    child{ node [visited] {}
            child{ node [visited] {} 
				child{ node [unvisited] {}}
				child{ node [unvisited] {}}
            }
            child{ node [visited] {}
				child{ node [unvisited] {}}
				child{ node [unvisited] {}}
            }
		}
; 
\end{tikzpicture}
\caption{An illustration of the breadth-first search (BFS) exploration strategy.}
\label{fig:BFS}
\end{figure}

Figure~\ref{fig:DFS} shows how a depth-first search (DFS) might look like. While this
exploration strategy results in very long paths, it also has some advantages. We need 
less memory to use DFS, since we can drop the states corresponding to already explored 
paths. Also, in breadth-first search (BFS, in Figure~\ref{fig:BFS}) the worklist size tends to be larger, which adds more extra memory. Moreover, using DFS, 
we are more likely to look up the same parts of the symbolic state during symbolic 
execution, which is also a great performance benefit.

\vspace{5pt}

\vspace{-5pt}

In order to combine the advantages of BFS and DFS, we can define \textit{analysis budgets}. 
These budgets can have a big impact on the shape of the built exploded graph, on run time 
performance, coverage, and the false positive ratio. The analyzer has limits such as the 
number of nodes to build for a top-level function, the maximal stack size during 
inlining, maximum number of times the same basic block on a path can be visited, and 
maximum size for a function to inline. Using these budget limits, we can enjoy the 
benefits of DFS while mitigating some of the drawbacks. For example, the analyzer will 
not end up spending all its time exploring one really long path, never visiting the 
rest of the code. These limits also prove to be useful while using other
exploration strategies, like the default unexplored-first method.

Exploration strategy has a profound impact on false negatives, as it might cause the 
engine to miss faulty code paths, but it is also important for the false positive ratio,
as the number of explored infeasible paths depends on it. One way to reduce 
the probability of visiting infeasible paths is to use \textit{trace-guided exploration}. 
During trace-guided exploration we record a set of paths from a real execution of the 
program (called \textit{traces}) and derive further paths from feasible traces by
either following a trace symbolically, or by generating a new trace from an 
old one, negating one of the path conditions. This method is mostly used in 
\textit{concolic execution}~\cite{sen2007concolic},
which stands for the combination of concrete and symbolic execution.
There are also other advanced exploration strategies to do
directed symbolic execution~\cite{ma2011directed} where the
tool prioritizes paths that can reach a certain program point.
In case we are interested in one particular kind of check it is
possible to create heuristics that guide the analysis towards paths
that are more likely to contain such errors \cite{kadar2017optimization}.

\subsection{Cross translation unit analysis (CTU)}

The scope of the analysis has a big impact on its precision. 
If we cannot reason about functions in separate translation
units, we will not find the errors that span across translation unit boundaries.
in Figure~\ref{fig:div_zero_false_neg} we see a division by zero error that
the analysis will not find unless it can reason about both \texttt{A.cpp} and
\texttt{B.cpp} together at the same time.

\vspace{-10pt}
\begin{figure}[H]
\noindent\begin{minipage}{.45\textwidth}
\begin{lstlisting}[caption=A.cpp,frame=tlrb]{Name}
int f(int x);

void g() {
  // Division by zero.
  // No warning below.
  int x = f(42);
}
\end{lstlisting}
\end{minipage}\hfill
\begin{minipage}{.45\textwidth}
\begin{lstlisting}[caption=B.cpp,frame=tlrb]{Name}
int f(int x) {
  // Potential
  // division by zero.
  return 5 / (x - 42);
}
\end{lstlisting}
\end{minipage}
\caption{Example of a false negative error spanning across translation unit boundaries.}
\label{fig:div_zero_false_neg}
\end{figure}

\vspace{-30pt}

\begin{figure}
\noindent\begin{minipage}{.45\textwidth}
\begin{lstlisting}[caption=A.cpp,frame=tlrb]{Name}
bool f(int x);

void g() {
  int x = 0;
  if (f(x)) ++x;
  // Warning below:
  // division by zero.
  int c = 5 / x;
}
\end{lstlisting}
\end{minipage}\hfill
\begin{minipage}{.45\textwidth}
\begin{lstlisting}[caption=B.cpp,frame=tlrb]{Name}
bool f(int x) {
  return x % 2 == 0;
}
\end{lstlisting}
\end{minipage}
\caption{Example of a false positive error spanning across translation unit boundaries.}
\label{fig:div_zero_false_pos}
\end{figure}
\vspace{-10pt}

Extending the scope of the analysis has other benefits in addition to finding issues that
span across translation units. We can also eliminate false positive findings.
In some cases, the analyzer cannot deduce that a path is infeasible due to a lack
of information that is only available in a separate translation unit.
Consider Figure~\ref{fig:div_zero_false_pos}. If the analyzer
knows the body of function \texttt{f}, it can deduce that the \texttt{if} branch is
always taken and there is no division by zero. If the body of the function
is unknown, the analyzer assumes that the branch might not be taken and reports a
division by zero error.

We can approach this problem in a multiple ways. Some projects pursue 
\emph{unity builds}~\cite{Mihalicza2010}, which is a way to create one big translation
unit that is fed into the compiler. This way it is easier to achieve whole program analysis
and also reduces compilation time. Unfortunately, it does not support incremental
compilation and parallelization of the build process. Moreover, right now there is no way
to automatically create a unity build from a project. The main advantage of unity build
is that we can use existing tools that only support analyzing a single translation unit
and use them for whole program analysis without modification.

In the upcoming C++ standard \emph{modules} will be introduced. They will mitigate
some problems with the compilation model, but will not eliminate them completely. The
introduced solution is backward compatible, but a significant amount of development
will be required to translate existing systems to use the new features.

An alternative way of extending the scope of the analysis is to use summaries, which
are described in Subsection \ref{summarybasedanalysis} in depth.

\begin{figure}
  \centering
  \includegraphics[width=.8\textwidth]{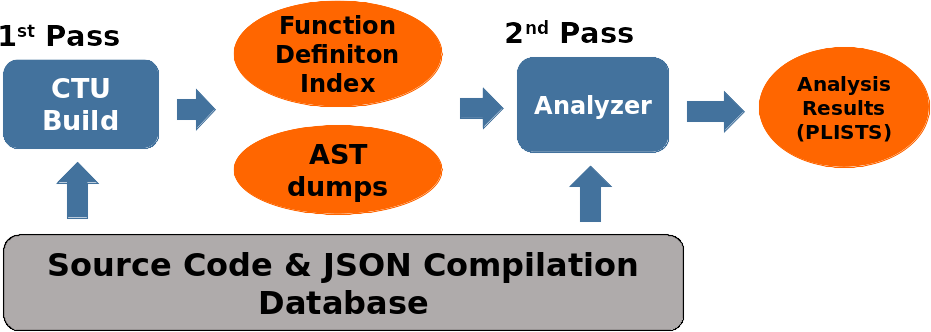}
  \caption{Two-pass cross translation unit analysis in the Clang Static Analyzer.}
  \label{fig:twopass}
\end{figure}
\vspace{-15pt}

\subsubsection{Architecture}
Originally, the Clang Static Analyzer analyzed each translation unit in complete 
separation, in one pass. To implement cross-translation unit analysis, we have extended
the analysis process with a \emph{second pass}~\cite{horvath2018engineering}. 
The architecture is shown in Figure~\ref{fig:twopass}. For C family languages, we need the
compilation command to be able to parse the source code properly. Both
passes require a database that contains the compilation command for each
translation unit. The first pass parses the translation units and builds up
an abstract syntax tree. 
We serialize ASTs to the disk in a binary format, and create an 
\emph{index file} that records function definition locations.

It is worth to mention that function names cannot be keys in the index file.
Function names are not unique in C++ due to several features such as templates,
overloading, and namespaces. We use a type of identifier called USR, which stands 
for \emph{Unified Symbol Resolution}.
This is a unique string that can be generated for each declaration by the
Clang compiler. The main purpose of its introduction into the compiler was
to support the \textit{go-to-definition} and similar functionalities of integrated
development environments (IDEs).

In the second pass, the analyzer will analyze each translation unit. When it
encounters a call to a function that does not have an implementation
in the current translation unit, the analyzer will look the function definition up in
the index. If the definition is in the index, the analyzer can locate
the corresponding AST dump and load its contents into memory. The result of
the analysis will be a \texttt{plist} file for each translation unit, which is
a form of XML that encodes dictionaries and arrays holding information about all
the errors in that unit.

\subsubsection{AST merging}

Even after loading the AST of the other translation unit holding the desired
function definition into memory, we are still not done. We have two separate ASTs, 
one representing the original translation unit that is being analyzed, and one 
representing the other translation unit that was loaded from the dump. 
These ASTs have separate symbol tables, separate representations of the same types, 
and their source locations are in separate file managers. 
We need to merge the two ASTs to be able to do the same inline analysis for calls 
\emph{across} translation units that we do \emph{within} a unit. 
In Clang there exists a module to merge ASTs called \texttt{ASTImporter}.
Unfortunately, before our work on cross translation unit analysis began, this 
module had been fairly incomplete. A large portion of our initial effort went into
adding C++ support to the AST merging process. 
Merging ASTs can also discover one definition rule violations, when the same
type has semantically different representation in multiple translation units.
We were able to find such cases in open source projects.

\subsubsection{Caching}
Both merging the AST and loading it from the disk can be expensive. So we
use two levels of \emph{caching}. If an external AST is requested for the second time
during analysis, we do not load it again, but keep it in memory.
Once we have already merged a subtree of the AST, we do not attempt to merge
it again.

\subsubsection{Consistency}
In the original analysis, an inlined function will not be re-analyzed as a top-level 
function. Should we implement similar behavior across translation units? 
For example, if a function has already been analyzed during a 
call chain originating from a separate translation unit, should we consider it as a top
level function, when we analyze the unit where its definition can
be found? The answer is yes. Usually, translation units are analyzed
in parallel using separate invocations of the analyzer. If the decision whether
we analyze a function as top-level depends on the analysis of other translation
units, we introduce a race condition. We can construct cases where the result of
the analysis depends on the order in which the units were processed.
It is important to have consistent results across runs, otherwise the users of
the analyzer cannot be sure whether they properly fixed the reported issue, or
the fluctuation of reports is due to the non-determinism in the analysis.

\subsubsection{Evaluation}
Surprisingly, mostly due to the well-defined budgets, it was feasible to run cross 
translation unit analysis on large projects using a personal computer. Run time and
memory consumption numbers can be inspected in Figure~\ref{fig:ctuperf} in Appendix B.
While the analysis did take longer and consumed more memory, the increase was 
proportional to the number of reports found. We also observed a slight improvement 
in the false positive ratio as seen in Figure~\ref{fig:fpratio} in Appendix B.
Also note that the shape of the exploded graph can change due to the increase in the 
scope of the analysis, resulting in a change of coverage patterns. Thus, while we do 
see new findings, we can also lose some of them (see Figure~\ref{fig:lostbugs} in 
Appendix B). Some of the findings might be lost due to being refuted by the 
additional information available to the analyzer while others are due to coverage changes.

\subsubsection{Future work}
The main setback of cross translation unit analysis is the lack of
support for consistent results when we use incremental analysis. The reason is that 
we not only need to recheck the changed files, but all the files that contain paths 
that enter the changed files. If we re-run the analysis for all those files, it will no 
longer feel incremental. Solving this issue is going to take a significant amount 
of research and engineering effort.

\subsection{Summary-based analysis}
\label{summarybasedanalysis}

Inlining is not the only way to implement inter-procedural analysis. An alternative is called
called \emph{summary-based analysis}, and it works by summarizing the semantics of a function and
using the summary to evaluate the effects of the function at the call site.
As the name suggests, these summaries will only estimate the behavior of the function.
These kind of analyses are never as precise as inlining but they tend to be more scalable.

Often, summaries are specific to a kind of analysis. For example, we could think of
the type signature of a function as a summary for type checking, which is also 
static analysis. For symbolic execution we encountered two approaches. The first one
uses \emph{logical formulae} as its abstraction and encodes information about the pre- and
post-states of a function call. Such summaries have the form
$\phi_1 \Rightarrow (\phi_2 \lor \phi_3) \land \phi_4 \Rightarrow \phi_5 ... $, where each 
$\phi$ is a predicate over the program state. The left side of the implication corresponds
to the state before the call, and the right side corresponds to the state after the call.
The other approach uses \emph{state machines} as its abstraction. An example 
is~\cite{xie2005scalable}.

There are certain classes of summaries. \emph{May summaries} are over-approxima\-tions.
The may summary $\phi_1 \Rightarrow \phi_2$ means that no states satisfying $\neg\phi_2$
are reachable from states satisfying $\phi_1$. \emph{Must summaries} are under-approximations.
The must summary $\phi_1 \Rightarrow \phi_2$ means any state satisfying $\phi_2$ is
reachable from any state satisfying $\phi_1$. There are methods to use both may and must
summaries to improve the analysis~\cite{godefroid2010compositional}.

Inferring preconditions for a function is a great way to generate summaries.
One well established approach is using \emph{backward symbolic execution}
~\cite{chandra2009snugglebug}.

Unfortunately, the analyzer does not implement general function summaries at the point of
writing this paper. It does have, however, a set of \emph{modeling checks} that act as summaries
for certain widely used functions like those in the C standard library. These checks
are written manually. While it is great to support hand written summaries, the goal is to
automatically generate them from implementations. Another mitigation for the lack of
summaries is the support for \emph{annotations}. There are certain annotations that greatly
improve the precision of the analysis, but we rely on the user to add them.
One such annotation is to mark functions that will never return, such as
assertions. This is essential to avoid false positives from infeasible paths.

\vspace{-5pt}
\begin{figure}
\begin{lstlisting}[frame=tlrb, language=xml]{Name}
<function name="strcpy">
  <leak-ignore/>
  <noreturn>false</noreturn>
  <arg nr="1">
    <not-null/>
  </arg>
  <arg nr="2">
    <not-null/>
    <not-uninit/>
    <strz/>
  </arg>
</function>
\end{lstlisting}
\caption{An example summary written for CppCheck.}
\label{fig:summary}
\end{figure}
\vspace{-5pt}

CppCheck also supports user-written summaries that capture information about
pre- and postconditions of a function. Figure~\ref{fig:summary} shows an example.
Further efforts exist to use textual representation for summaries~\cite{Horvath:2016}.
The basic idea is to write code that under- or over-approximates the
original function using the same language.

\subsubsection{Modeling loops}
Functions are not the only entities that can be summarized. Some analyses are attempting
to summarize the effects of \emph{loops}~\cite{godefroid2011automatic}.

Right now the analyzer uses a very primitive method to model loops. We \emph{unroll}
loops up to three times to avoid too many iterations. Each iteration might split
paths multiple times, thus we might end up having an explosion in the number of paths
to explore. This under-approximation helps us keep the run time reasonable and
the number of false positives low. A recent addition to the analyzer detects loops of a specific
form that will not split paths too often. Those loops are unrolled completely.

One problem is that we might use up all our budgets during unrolling a loop, resulting in
a coverage loss. In order to regain that coverage, we implemented \emph{loop widening},
which over-approximates the effects of the loop and continues the execution after
the loop. While it improves coverage, we need to do a significant amount of
invalidation for the over-approximation, which can lead to additional false positives.
For this reason, this feature is off by default. One reason it needs to do an 
excessive amount of invalidation is the lack of pointer analysis. Since we
do not execute the loop body symbolically, we need some other oracle to
get the possible pointees of each pointer and reference to know which values to
invalidate. An abstract interpretation method is a viable candidate and we do plan
to implement it in the future.

\subsection{Checkers}

This paper is not intended to be a guide on how to write a \emph{check} 
(or \emph{checker}) for the Clang Static Analyzer, but we 
do wish to give a brief idea about how they interact with the engine. In case you want to develop new
checks we recommend the following guides:~\cite{annazaks,noq,devguide}.

Checks in the analyzer can be classified into two main categories. One comprises 
\emph{modeling checks} that will not report any warning, but instead help the engine simulate the 
program. Modeling checks represent domain-specific knowledge about a codebase, e.g. simulate the behavior 
of standard library functions like \texttt{toupper}. Checks belonging to the other class \emph{find errors}.

Each check can subscribe to a set of events like function calls or division operations, and
maintain their own state for each path. Checks can also query the state, but instead of dealing with
specific symbolic expressions, authors are advised to ask questions like \emph{Can the second argument
of this function be zero on this path?} This query will be forwarded to an SMT solver.

\subsubsection{Statistical checks}
The classical approach of writing checks is to encapsulate some \emph{domain-specific knowledge}
about the expected
behavior of the program. For example, we expect each memory allocation function call to be paired with
a memory deallocation function call on each path. The problem of this approach is that we need to know
which functions allocate or deallocate, but lots of system programs are using non-standard APIs for these
purposes. An alternative approach is to assume that most of the code is bug free and try to infer
the rules from the source code using statistical methods. A prominent method is described
in~\cite{kremenek2006uncertainty}.

The upstream version of the Clang Static Analyzer does not support statistical analysis, as it does 
not support arbitrary passes. Usually, we do not want to restrict statistical inference to a single 
translation unit a time, but rather run it as a separate pass over the whole source code before analysis.
We have, however, a fork of Clang
which supports some basic statistical checks that infer annotations 
such as which function's return values are required to be checked. If a function's return value is 
checked 99\% of the time, it is very likely that the rest of the calls with unchecked return values 
are not correct. We do plan to upstream this infrastructure in the near future.

\subsubsection{Bug reports}
While the analyzer will show us the path leading to each reported bug, it can still be challenging to 
understand the reasons behind a certain report. Thus, checks can add additional notes or events to
execution paths that are rendered for the user. For instance, for a memory leak error, the check can
add a note indicating the point where the leaked memory was allocated. This is a great help in deciding 
whether a warning is a false positive or a real error.

A closely connected concept is that of the \texttt{Bug Reporter Visitor}, which traverses the execution 
path leading to the bug, adding notes before they are displayed to the user. Built-in visitors 
exist that track the source of certain values. This can help display information like where the zero came 
from in case of a division by zero error.

\subsubsection{False positive suppression}
A visitor can also mark a bug report false positive. Those marked errors will not be presented to the user.
One of the visitors filtering false positives is based on so-called \emph{inline defensive checks}. 
Let us consider the code on Listing~\ref{lst:inline-defensive}:

\vspace{10pt}
\begin{lstlisting}[frame=tlrb, caption={A warning filtered out by the \textit{inline defensive check} heuristic.}, captionpos=b, label={lst:inline-defensive}]
int f(int *p) {
    if (p == 0)
        return 0;
    // ...
    return x;
}

void g(int *q) {
    int a = f(q);
    int b = *q; // Should we warn?
    // ...
}
\end{lstlisting}
\vspace{5pt}

\noindent Should we warn about a possible \texttt{null} dereference of \texttt{q} in function \texttt{g}?
The analyzer \emph{would} warn if we disabled the visitor mentioned before, because there exists
a path where function \texttt{f} returns \texttt{0} and the value of \texttt{q} is also
\texttt{0}. On this path we will dereference \texttt{q}. The check in \texttt{f}, however, might
be redundant for \texttt{g}, if \texttt{g}'s author knows that in \texttt{g}'s context 
\texttt{q} may never be \texttt{null}. In other words, we do not know for which set of callers 
this check is relevant, so it is safer not to warn. 
The inline defensive check visitor will prevent issuing such warnings.

\vspace{10pt}
\begin{lstlisting}[frame=tlrb, caption={A warning that is \textbf{not} filtered out by the \textit{inline 
defensive check} heuristic.}, captionpos=b, label={lst:inline-defensive-not}]
void g(int *q) {
    int a = f(q);
    if (q == 0)
        a = 0;
    int b = *q; // Should we warn?
    // ...
}
\end{lstlisting}
\vspace{5pt}

\noindent For the code on Listing~\ref{lst:inline-defensive-not}, the analyzer will warn
regardless of the warning suppressing heuristics, since
the developer definitely considered the case of \texttt{q} being null. Thus we can assume that \texttt{null}
is a valid input for \texttt{g}. So it is safe to warn for the possible \texttt{null} dereference.

As we can see, infeasible paths are one of the biggest challenges of symbolic execution.
Since the analyzer is using a fast but simple solver, it will not attempt to warn when it
reaches an assertion failure. Instead, it will conclude that the execution path was infeasible.

There exist tools that are able to generate test cases based on the results of symbolic 
execution. The Clang Static Analyzer did not pursue this direction because of the simplicity
of the default constraint solver. If the analyzer ends up emitting a test case that does 
not trigger the behavior, the user might have a false sense of security, even though it 
might still be possible to trigger it with a different test case.

\subsection{Refutation}

Despite heavy developer effort, the analyzer suffers from the problem of
false positive reports much like other similar tools. One possible way to improve
report quality is to improve the constraint management of symbolic expressions, 
which plays an important role in proving the infeasibility of impossible
execution paths. An important intermediary step in this direction 
is the refutation of false positive reports by re-evaluating constraints by a more 
powerful constraint solver than the one currently built into the engine~\cite{tiered}.

\paragraph{The range-based constraint solver}
The analyzer collects constraints on symbolic variables encountered 
in the program to be able to detect if they become unsatisfiable. Solving these constraints is 
only one side of the coin: generating and managing them is another. Support for constraint 
management is therefore scattered throughout the analyzer engine. The current solution centers 
around a solver operating on range-based constraints, which is only capable of handling some 
common binary operations between symbolic values and concrete integers (called 
\texttt{SymIntExpr}s), and some relational operations between two symbols (\texttt{SymSymExpr}s). 
Although it is very fast, it lacks support for many other commonly used arithmetic operations 
even on \texttt{SymIntExpr}s, such as bitwise operations, multiplication, division, etc.

\paragraph{A more powerful solver}
In 2017, support for an alternative constraint solver backend, the Z3 Theorem Prover,
has been added to the engine. Z3 is a state-of-the-art general-purpose SMT solver developed 
by Microsoft Research. It is capable of handling most arithmetic operations not supported by 
the current solver, such as those on floating-point values, and it also represents integers 
more realistically, modeling them with fixed-width bitvectors.

Unfortunately, the analyzer will not be able to harness the full power of Z3 until 
symbolic expression support is improved in the engine. The analyzer currently does not 
build up symbolic expressions consisting of floating-point type values, and subsequently does 
not generate constraints on them, meaning that information about such expressions never arrives 
at the constraint manager. Still, without any further effort, Z3 should already 
be able to improve analysis precision for expressions involving pointers and integers.

Nevertheless, the analyzer still does not employ Z3 as the default constraint solver backend. 
The reason behind this is its negative impact on the duration of the analysis, with execution 
times soaring up to and above a factor of 20 times the usual. This slow-down stems from the 
nature of SMT solvers, which follow complex inner heuristics, and often use up all of the 
allowed time as limited by the timeout parameter for a single operation. For practical use, 
an intermediary solution is needed.

\subsubsection{Experimental comparison of the two solvers}
In an effort to explore how each of the currently available constraint solving backends 
affect analysis performance and quality, we made the following experiment. For 3 real-world 
open-source projects, we ran two analyses, each with default settings but differing in the 
use of the constraint manager backend. We were concerned about the number of reports and 
execution times in each case. In Table~\ref{tab:z3}, the RB keyword denotes the default 
range-based solver built into the engine, while reports added and removed are meant for 
the Z3 cases compared to the runs using the range-based solver. Analysis duration
is showed in \texttt{hh:mm:ss} format.

\vspace{-5pt}
\begin{table}[h!]
\centering
\begin{tabular}{|c|cccccc|}
\hline
\shortstack{Project\\name} & \shortstack{Reports\\(RB)} & \shortstack{Reports\\(Z3)} & \shortstack{Reports\\removed} & \shortstack{Reports\\added} & \shortstack{Duration\\(RB)} & \shortstack{Duration\\(Z3)} \\ \hline
tmux & 19 & 19 & 0 & 0 & 00:01:07 & 03:03:10 \\
redis & 243 & 64 & 185 & 6 & 00:03:04 & 05:42:53 \\
xerces-c & 88 & 2 & 86 & 0 & 00:04:59 & 01:41:32 \\ \hline
\end{tabular}
\vspace{0.2cm}
\caption{Report counts and run time durations of default-configured analyses run using 
either the range-based (RB) or the Z3 constraint manager back-ends on 3 open-source 
projects. The \emph{Reports added} and \emph{Reports removed} columns compare Z3 against the
range-based solver.}
\label{tab:z3}
\end{table}

\vspace{-20pt}

We can observe that using Z3 as the constraint management system throughout the whole
analysis process elongates run time from a matter of minutes to a matter of hours. As the
engine builds the exploded graph data structure for each analyzed top-level function in a
project, it keeps requesting operations from the Z3 back-end that are more expensive than 
those of the range-based back-end. For a million source files, this adds up to a run time
that is no more feasible for industrial use.

\subsubsection{Refutation in action}
One possible compromise is to use the Z3 back-end for \emph{false positive refutation}. 
This means to perform the analysis as usual, then post-process the collected bug reports and 
remove those that lie on paths that are found to be infeasible by Z3. Thus, instead of 
participating in the analysis of a million source files, Z3 is only presented with at most
a few hundred bug reports that need to be cross-checked. This method can eliminate a large 
portion of false positive reports while only introducing a tolerable burden on the run 
time of the analysis.

In order to see how the use of Z3 for refutation can help us remove false positives,
consider the example on Listing~\ref{lst:refut-example}.

\vspace{10pt}
\begin{lstlisting}[frame=tlrb, caption={A sample function causing the analyzer to emit
a false positive warning on line 8, due to insufficient constraint handling.}, captionpos=b, 
label={lst:refut-example}]
void g(int d);
void f(int *a, int *b) {
  int c = 5;
  if ((a - b) == 0)
    c = 0;
  if (a != b)
    g(3 / c); // Warning: division by zero.
}
\end{lstlisting}
\vspace{5pt}

\noindent Arriving at the second \texttt{if}, both conditions are understood and translated to 
ranged constraints, but the solver is not able to prove that they contradict each other. This 
can be seen from the exploded graph as the current path splits to two, meaning that the constraint 
manager found both new states to be feasible. On the path assuming that both conditions are true, 
the exploded node holds the constraints that can be seen on 
Listing~\ref{lst:refut-example-constraints}:

\vspace{10pt}
\begin{lstlisting}[frame=tlrb, caption={Range constraints collected for pointers \texttt{a} 
and \texttt{b} during the analysis of function \texttt{f} on Listing~\ref{lst:refut-example},
up until line 8.}, captionpos=b, label={lst:refut-example-constraints}]
Ranges of symbol values:
(reg_$0<int * a>) - (reg_$1<int * b>): { [0, 0] }
(reg_$1<int * b>) - (reg_$0<int * a>): { [PTR_MIN, -1],
                                         [1, PTR_MAX] }
\end{lstlisting}
\vspace{5pt}

\noindent Here, the \texttt{a != b} condition has been rearranged by the engine to \texttt{(b - a) != 0)},
and the constraint was generated by substracting zero from the full range of possible pointer
values, representing the result with a union of the two intervals below and above zero.
\noindent These constraints can be modeled by the small \texttt{z3} program on
Listing~\ref{lst:refut-example-z3}:

\vspace{10pt}
\begin{lstlisting}[frame=tlrb, language=Lisp, caption={A small program written in Z3's native \texttt{SMT-LIB}
format, encoding the range constraints seen on Listing~\ref{lst:refut-example-constraints},
and checking their satisfiability. The \texttt{unsat} result it gives means that the path
on which these constraints have been collected is infeasible.}, captionpos=b, label={lst:refut-example-z3}]
(declare-const a (_ BitVec 32))
(declare-const b (_ BitVec 32))
(assert (= (bvsub a b) #x00000000))
(assert (bvslt (bvsub b a) #x00000000))
(assert (bvsgt (bvsub b a) #x00000000))
(check-sat)
(get-model)
\end{lstlisting}
\vspace{5pt}

\noindent After execution, the program on Listing~\ref{lst:refut-example-z3} gives an 
\texttt{unsat} result, i.e. the problem is proved to be unsatisfiable. This is the simplest 
case refutation should be able to handle: ranged constraints are readily available in the 
program state, they only need to be fed to a Z3 solver instance in the proper format. 
For this, constraints on symbolic values need to be converted to the internal expression type 
used by Z3, which involves the translation of integer relations into their corresponding 
correct bitvector operations. If the translation succeeds and Z3 can prove the state to be 
infeasible, the report is marked invalid, and never shown to the user.

\subsubsection{Evaluation}

The false positive refutation option was designed to provide a compromise between the speed 
of the default analysis and the precision of an analysis using the Z3 constraint manager backend. 

We cannot expect the same results for several reasons. First, analyzing projects using the Z3
back-end, the whole process uses the Z3 constraint manager, and the resulting exploded graph may 
differ from the one built in default mode. This means that constraints stored in the graph may 
be slightly more realistic or precise than those generated in the default mode. However, its 
working mechanism also differs from the case in which the default analysis is merely enhanced by 
the refutation visitor. Because of its independent nature, refutation may eliminate false reports 
that an analysis with the Z3 back-end cannot, e.g. those caused by weaknesses in the engine's 
general operation. Results are shown in Table~\ref{tab:ref1}.

\vspace{-5pt}
\begin{table}[h!]
\centering
\begin{tabular}{|c|cccccc|}
\hline
\shortstack{Project\\name} & \shortstack{Reports\\(RB)} & \shortstack{Reports\\(REF)} & \shortstack{Reports\\(Z3)} & \shortstack{Duration\\(RB)} & \shortstack{Duration\\(REF)} & \shortstack{Duration\\(Z3)} \\ \hline
tmux & 20 & 16 & 19 & 00:01:01 & 00:01:18 & 03:03:10 \\
redis & 243 & 161 & 64 &  00:02:15 & 00:04:01 & 05:42:53 \\
xerces-c & 88 & 40 & 2 &  00:03:22 & 01:01:22 & 01:41:32 \\ \hline
\end{tabular}
\vspace{0.2cm}
\caption{Comparison of analyses run with the default configuration (RB), with refutation 
enabled (REF) and using the Z3 constraint manager back-end (Z3).}
\label{tab:ref1}
\end{table}
\vspace{-20pt}

Next, we compared the performance of a default analysis to one with refutation enabled, but
on more projects, and weighing whether the number of reports removed justifies any run time
penalties, as essentially this was our most important concern. Table~\ref{tab:ref2} 
contains the number of bug reports for two analysis runs for 6 open-source projects, one with 
a default configuration, and one with the naive prototype of false positive refutation enabled.

\begin{table}[t!]
\centering
\label{tab:ev}
\begin{tabular}{|c|ccccc|}
\hline
\shortstack{Project\\name} & \shortstack{Reports\\(RB)} & \shortstack{Reports\\(REF)} & \shortstack{Reports\\removed} & \shortstack{Duration\\(RB)} & \shortstack{Duration\\(REF)} \\\hline
tmux & 20 & 16 & 4 &  00:01:01 & 00:01:18 \\
redis & 243 & 161 & 82 &  00:02:15 & 00:04:01 \\
xerces-c & 88 & 40 & 48 &  00:03:22 & 01:01:22 \\ 
libWebM & 32 & 28 & 4 &  00:01:21 & 00:02:50 \\ 
curl & 42 & 36 & 6 &  00:01:01 & 00:01:00 \\
bitcoin & 871 & 865 & 6  & 00:29:30 & 00:40:17 \\ \hline
\end{tabular}
\vspace{0.2cm}
\caption{Report numbers for analyses with a default configuration (RB) and with false positive 
refutation (REF) enabled for some open-source projects.}
\label{tab:ref2}
\vspace{-15pt}
\end{table}

Both the number of invalidated reports and the difference in analysis duration depends heavily
on the analyzed project. For example, \texttt{xerces} seems to have needed a lot of 
post-processing work, with more than a half (55\%) of its reports removed and experiencing an 
excessive increase in execution time (by a factor of 20). On the other hand, no temporal 
overhead of false positive refutation can be observed in the case of \texttt{curl}. After 
performing a manual inspection of some of the removed reports, either the falseness of the 
reports was difficult to determine (because of long bug paths), or we found that the report 
was truly a mistake on the analyzer's behalf and its removal increased the overall quality 
of the analysis.

One disadvantage of using refutation is the existence of the timeout parameter.
It is no longer deterministic whether Z3 can refute a bug or not so the displayed
results can oscillate between runs. While this could certainly happen we never observed
it.

\subsubsection{Related work}
There were attempts to use other solvers for refutation besides Z3~\cite{mikhail}.
Also, there are experiments to use machine learning to fine tune 
solvers~\cite{balunovic2018learning}
for a certain use-case. While it is an interesting idea we never experimented with this
direction. An alternative approach to reducing the run time of analysis with Z3 would be to
only use Z3 for certain some decision but not all. It is possible to come up with some
heuristics based on the number of constraints on the symbols when should we query to
more sophisticated solver. This is something that we might experiment with in the future.

\section{Tooling}
\label{tooling}

We define the scalability of static analysis not only in terms of efficient
use of computing resources, but also in terms of efficient use of
human resources like developer time\footnote{Based on the Intellectual output 
O1 part/topic ``Static Code Analysis with CodeChecker'' of project No. 
2017-1-SK01-KA203-035402 (for more details see Acknowledgement).}.  
CodeChecker~\cite{cc} is a tool designed to ease the 
integration of the analyzer and other similar static analysis tools into build 
systems and continuous integration loops. It is also a full-fledged bug management 
system that keeps track of errors found by these tools.

\subsection{Build systems}

Different static analysis tools have different strengths and weaknesses. Thus it is 
often advised to use multiple tools. One problem of using multiple tools is that
we need to integrate all of them into the build system and the continuous integration
loop. Unfortunately, this is not a one time cost, since these build systems and CI
loops tend to change over the course of a project, having these tools integrated
can add a maintenance cost later on. One purpose of CodeChecker is to reduce the burden on the
developers and make the experience more streamlined. CodeChecker supports multiple
static analysis engines and we plan to add more in the future. If a project is using
CodeChecker, they only need to integrate one tool into their workflow, which
was designed with integrators in mind from the start, and they
get multiple analyzers for a constant cost (in the number of analyzers).

The method we found most robust for integration is to capture compiler
invocations issued by the build system by hijacking the \texttt{exec} system call family.
Each time the build system invokes an external binary, CodeChecker will be
notified and it can log the build command.
We also support incremental analysis, because we rely on the original 
build systems to decide which files are compiled (thus need to be re-analyzed).
Another advantage of being plugged into the original build system is supporting
code generators. If a build step generates code before compilation, 
CodeChecker will see the generated code. 

\begin{figure*}
\includegraphics[width=\textwidth]{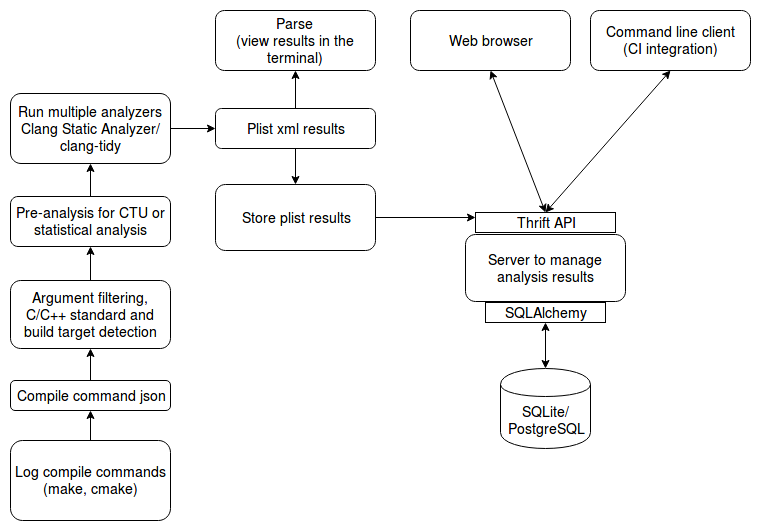}
\caption{CodeChecker architecture.}
\label{fig:arch}
\end{figure*}

The architecture of CodeChecker can be seen in Figure \ref{fig:arch}. As you can see our
job is not finished after having logged the compilation commands. Different compilers
use different flags. We added a layer to translate some of the flags from the
most common compilers like GCC to Clang. CodeChecker also supports multi-pass
analysis. Results are stored into a database and a web server will
serve  clients through a remote procedure call interface.

\subsection{Presenting the warnings}

Given a finite budget of developer time and thousands of reports on large software, 
it is important to evaluate reports with the best return on investment first.
CodeChecker helps prioritize easy-to-evaluate and high-severity bugs for developers,
aided by advanced filtering capabilities that find reports that are likely to be relevant to 
solving a certain problem. 

\begin{figure}
\includegraphics[width=\textwidth]{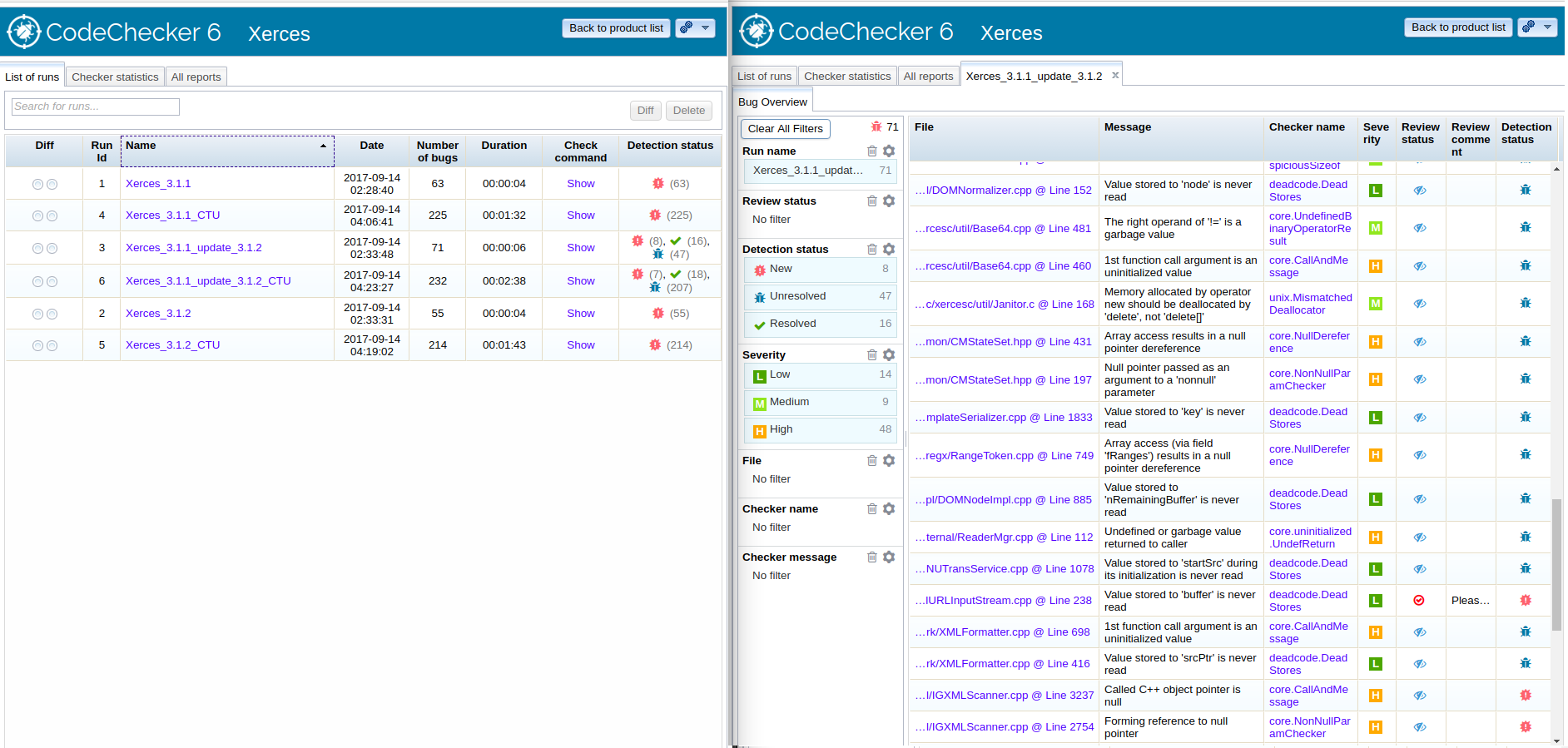}
\caption{CodeChecker's bug list view allows the user to browse and filter warnings.}
\label{fig:buglist}
\end{figure}

The bug list view showcasing some of the filtering capabilities can be seen 
in Figure~\ref{fig:buglist}.
The current version is using built-in heuristics for \emph{ranking}, but in the future we
plan to introduce machine learning based methods. The idea is not new, 
other researchers already reported a significant increase in static analysis tool
usability by introducing such rankings~\cite{kremenekstatic,kremenekdynamic}.
One of the challenges is to extract all useful features from the analysis 
engine, as it was originally designed to emit only the information necessary
to understand the bug reports.

For example, the Clang Static Analyzer does not support computing a \emph{confidence value}
for each
bug report. It would be a great help for ranking systems. While at first seems to be an
easy task to use a probabilistic approach reducing the confidence each time we cannot
decide which branch should we take on a certain path, the main challenge is how to 
increase the confidence when the same bug is found on multiple paths. We cannot just
add the probabilities for each path since these paths might not be completely independent.

CodeChecker also supports \emph{differential analysis} that prevents developers from
introducing new bugs without requiring them to fix all legacy reports beforehand.
This is a useful feature, because developers in the industry are often reluctant to
change old but well-tested code. And this is very reasonable since new code is
more likely to have true positive results than well tested code.

\begin{figure}
\includegraphics[width=\textwidth]{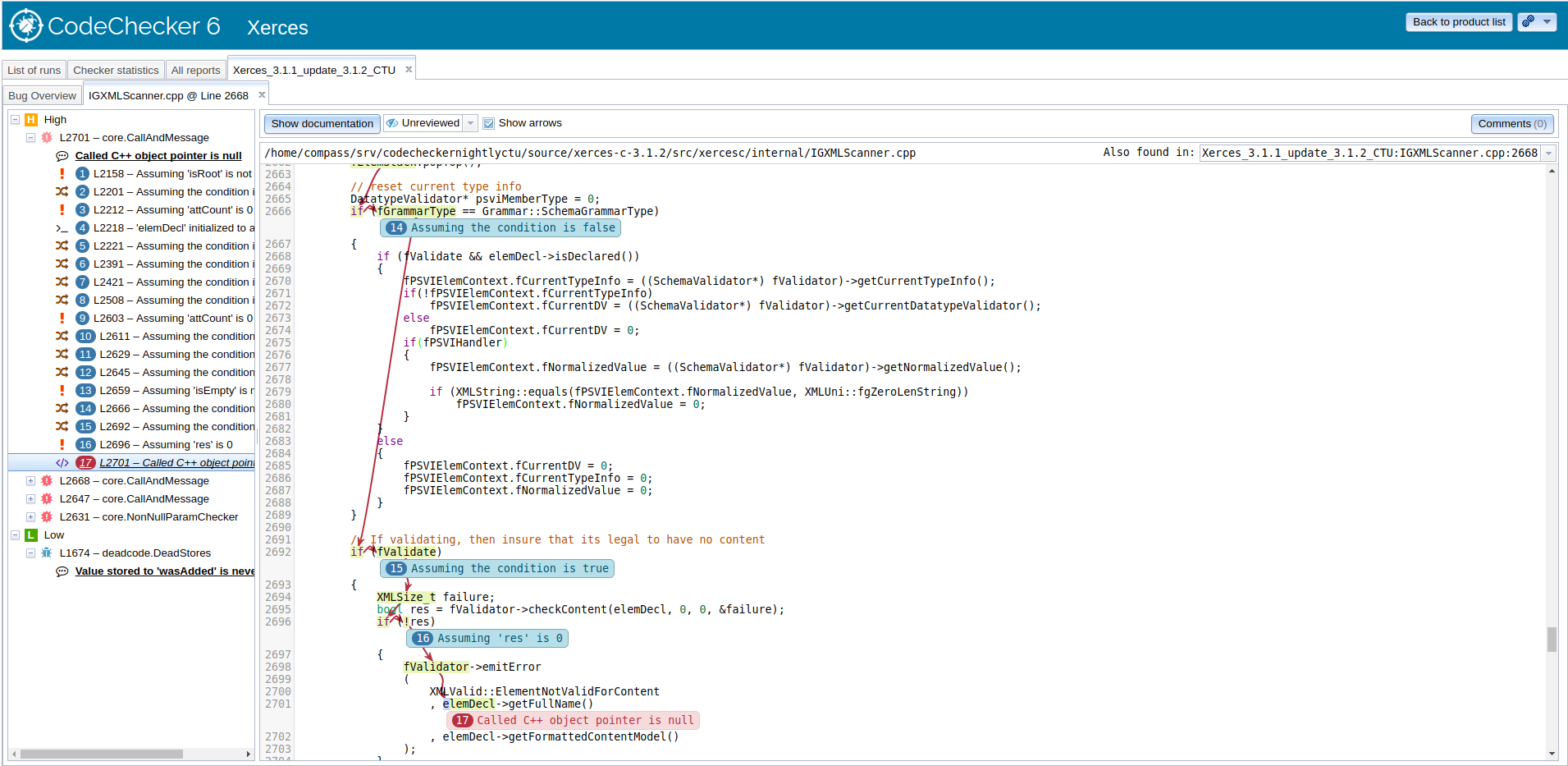}
\caption{A screenshot of CodeChecker's bug viewer window, which enables users to follow
through the execution path leading to a bug, highlighting important points along the path.}
\label{fig:viewer}
\end{figure}

One of the main advantages of CodeChecker is that it is able to display the path
emitted by the analysis tools (see Figure~\ref{fig:viewer}).
This way the developer will have a clue about how to reproduce a problem. 
We can also display important events along the path
such as where the object we forgot to release had been allocated. It is also
important to display the assumptions made by the analyzer, as it can aid both
the understanding of the problem and the classification of true and false positives. 

An unsolved challenge comes from the diversity of static analysis tools. The same
error might be found by different tools and all of them might represent it in a different way.
For example, they might find the same error on a different path, they might have separate
warning messages, they might even report the same error to different locations. 
It would be nice to present each error only once to the user regardless of how many tools
reported it. We do plan to add some heuristics to match these warnings.

In modern software development, communication is an important factor. The chances are good
that a developer might want to ask the opinion of a colleague when evaluating a report.
We added several features to support communication. We provide permanent links for each
report so that users can include those links in chat messages, emails or bug tracking
tools. We also let users comment to issues, so that they can express their reasons for 
classifying a report as a true or false positive.

During the analysis of a C++ program, the analyzer might see a large amount of code that is
irrelevant to the user, for example third party headers. CodeChecker supports configuration
options to automatically exclude those results. By excluding those headers, CodeChecker will
not even store those reports, saving disk space for the user.

In an ever-changing software project it is always challenging to keep bug classification
up to date. In case of a false positive division by zero error, how do we match
the classification made for an older version of the software, to a newer one?
Developers might have moved the function containing the error into a new file, and 
added or removed multiple lines of code around the warning. We support two methods 
for suppressing false
positives permanently that are resilient to changes. The first one is using comments
near the warning, the second one is using hashes of the context of the warning.
We advise users to rely on comments and only fall back to hashes when they cannot
add comments, for example when the file that needs to be modified is maintained by another team.

Unfortunately, some analysis tools might also output some redundant information as part
of the path. While this problem needs to be fixed in the analyzer rather than in CodeChecker,
we will consider using static program slicing~\cite{slicing} for this purpose.

\section{Testing with the CSA Testbench}
\label{testing}

Static analysis engines are software just like a text editor or a browser. We need
to test them to make sure they work well and do not regress later on. Unfortunately,
testing and debugging static analysis related software is often a challenging problem.
One of the reasons is that the developer does not only need to deal with the source code
of the analysis engine, but also the source code under analysis.

The proposition of a new patch to a static analysis engine involves discussing the
possible effects of the change. For this, we normally need analysis results on quite a 
few software projects before and after applying the patch. 
Obtaining such results can be surprisingly cumbersome. First, finding a set of test 
projects that truly show the effects of the patch can be a challenging task. Second, 
a request to extend the number of test projects might result in a significant amount of 
extra work for the patch author. This extra work comes from different sources. 
With the continuous evolution of the static analysis engine, the author needs to re-run 
the analysis on the requested projects using the most recent version of the engine. 
The results also need to be processed, so that they can be easily digested by the 
reviewers. Ideally, reproduction should be painless, and it should be possible to 
present results in an easily shareable and digestible format.

In this section, we present a toolchain~\cite{csa-repo,testbench} for the Clang Static 
Analyzer (CSA), that we call \emph{CSA testbench}, which aims to improve the situation 
by supporting both reviewers and authors in the following ways:

\begin{itemize}
\item[\textbullet] help authors select a set of relevant projects for testing, and run static ana\-ly\-sis on them,
\item[\textbullet] aggregate statistics about the analysis (e.g.: how often a cut heuristic is triggered when building the symbolic execution graph),
\item[\textbullet] aggregate the results of the analysis,
\item[\textbullet] help authors and reviewers evaluate and share the results,
\item[\textbullet] help reviewers reproduce results and maintain tests.
\end{itemize}

The input of the toolset is a single configuration file.
The output is an HTML report with useful information and figures, which
also contains the input configuration for easier reproducibility.

\subsubsection{Semi-automatic test suite generation}

The conventional approach to testing the analysis engine is to run it on a number of projects.
Finding a sufficient amount of relevant real-world projects can be challenging.
Ideal projects should be open-source for reproducibility and should
exercise the right parts of the analyzer. For example, if the change is 
related to the modeling of dynamic type information, only projects
using dynamic type information should be included. One possible option is to
check a random sample of open-source projects, hoping to find enough of them that displays
all of the required traits. A slightly better approach is
to use code searching and indexing services and search for projects with
interesting code snippets. These services, however, are optimized to present
the individual snippets and suboptimal to retrieve the most relevant projects
according to some criteria.

To mitigate this problem, we created a script to harvest the results from
an existing code search service (\texttt{searchcode.com}) and to recommend projects to be included in the test suite based on the results.

For example, in order to test a new static analysis check written to 
detect \texttt{pthread\_mutex\_t} abuse, we might be interested in C and C++ projects 
that use \texttt{pthread} extensively. Using the syntax seen on 
Listing~\ref{lst:gen-project}, we can specify the keywords to search for, the 
languages we are interested in, the desired number of projects:

\vspace{10pt}
\begin{lstlisting}[frame=tlrb, language=bash, caption={A sample invocation of the project list
generator tool.}, captionpos=b, label={lst:gen-project}]
 $ ./gen_project_list.py 'pthread' 'C C++' 5 -o pthread.json
\end{lstlisting}
\vspace{5pt}

\noindent The call on Listing~\ref{lst:gen-project} creates a configuration file with the suggested projects in the JSON format showed on Listing~\ref{lst:config}.

\vspace{10pt}
\begin{lstlisting}[frame=tlrb, caption={A fraction of a configuration file generated
by the project list generator tool invocation showed on Listing~\ref{lst:gen-project}.}, captionpos=b, label={lst:config}]
{
  "projects": [
    { "url":"github.com/itkovian/torque.git",
      "name": "torque" }, ...
  ]
}
\end{lstlisting}

\subsubsection{Easy reproduction and sharing}

The next problem is sharing results with reviewers. A regular
pattern we see is that the patch author shares text files containing analysis
results on certain projects. However, text dumps of static analysis results are hard 
to interpret and the measurements are hard to reproduce. How did the author compile 
the project? Which version of the analyzed project was used? How did the author 
invoke the analyzer? What configuration options were used? What revision (commit) 
of the analyzer was used?

\begin{figure}[h!]
\includegraphics[width=\textwidth]{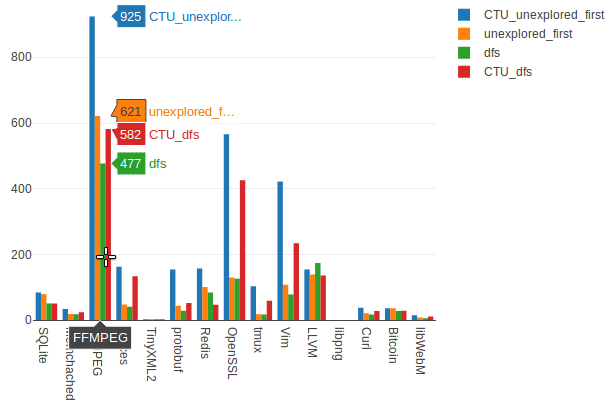}
\caption{An interactive chart generated by the CSA Testbench scripts.}
\label{fig:csa-chart}
\end{figure}

Our scripts use a concise configuration format that contains all the
relevant information about the analyzed projects: repository, tag/commit, configuration
options for the analysis, etc. Obtaining this configuration file enables
reviewers to reproduce the exact same measurements at their convenience. They can also 
easily suggest modifications to the conducted experiment.
Moreover, the results are not mere text dumps anymore but are presented on a convenient web
user interface that also displays the path associated with the report.
Other information such as the number of code lines of the project, version of
the ana\-ly\-zer, analysis time, analysis coverage,
and statistics from the analysis engine is recorded
and figures like charts are generated automatically.
An example figure from an HTML report is in Figure~\ref{fig:csa-chart}.

The file with the project list is almost enough to run the analysis on its own.
The only extra information needed to be specified is the CodeChecker server where
analysis results are intended to be stored for later inspection, as seen on
Listing~\ref{lst:server}.

\vspace{10pt}
\begin{lstlisting}[frame=tlrb, caption={A segment of the configuration file
specifying the address of the CodeChecker server.}, captionpos=b, label={lst:server}]
{
  "projects": ... ,
  "CodeChecker": {
    "url" : "localhost:15010/Default"
  }
}
\end{lstlisting}
\vspace{5pt}

\noindent After this we are ready to run the analysis on the previously selected set of 
projects, as shown on Listing~\ref{lst:experiment}.

\vspace{10pt}
\begin{lstlisting}[frame=tlrb, caption={Sample invocation of the main driver script
of the experiment.}, captionpos=b, label={lst:experiment}]
 $ ./run_experiments.py --config pthread.json
\end{lstlisting}
\vspace{5pt}

The script downloads each of the projects, attempts to infer their build system and 
build them, runs the analysis, and finally collects the results. At the time of writing 
this paper \texttt{autotools}, \texttt{CMake}, and \texttt{make} are supported as 
build systems.

In case a special build command is required or the build system is not yet supported,
the user can specify the build command. Building a special version of the project
characterized by a tag or a commit hash instead of top of tree is also possible
and highly encouraged to get consistent results with subsequent experiments.
Finally, differential analysis can currently be conducted by running the same
projects multiple times with different options passed to the analyzer or using
different versions of the analyzer. An example can be seen on Listing~\ref{lst:configs}.

\vfill
 \pagebreak

\vspace{10pt}
\begin{lstlisting}[frame=tlrb, caption={Differential testing can be achieved
by running many analysises on the same projects with different options passed 
to the analyzer.}, captionpos=b, label={lst:configs}]
{
  "projects": [ ... ], 
  "configurations": [
      {
        "name": "original",
        "clang_sa_args": "",
      },
      {
        "name": "variant A",
        "clang_sa_args": "argument to enable feature A",
        "clang_path": "path to clang variant"
      }
    ],
  ...
}
\end{lstlisting}
\vspace{-5pt}

\subsubsection{More precise differential analysis}

Finally, the number of tools available to support differential analysis on a project is scarce.
In case of the Clang Static Analyzer, we can only compare the number of bugs found,
analyzer engine statistics, and coverage percentage measured in basic blocks.
All of these are aggregated scalar values missing positional information,
with the statistics and the coverage being displayed individually for each translation unit.

We implemented analysis coverage measurement right in the heart of the analyzer
based on the \texttt{gcov} format. We do not calculate coverage as an overall percentage
value but record it separately for each line. This makes it possible to precisely aggregate
coverage information over translation units. This also makes it possible to do differential
analysis on the coverage itself. Our toolset includes scripts to aid that kind of analysis. 

In some cases, we are interested in the reason behind a specific bug report 
disappearing when running the analysis with different parameters. Performing 
differential analysis on the coverage, we are able to determine whether the 
analyzer actually examined the code in question during both runs.

The analyzer can output different kinds of statistics such as the number of paths
examined, the number of times a specific cut heuristic was used etc. Instead of having 
a fixed set of statistics to collect we used some heuristics to process the output of 
the analyzer, in which we are able to automatically detect statistics and aggregate 
them over translation units.

We create an HTML summary of the statistics collected during the analysis.
This report includes charts and histograms. We mine the log emitted by the analyzer 
to automatically collect all the statistics. After adding a new statistic
to the analyzer engine the author only needs to add a single entry in the
configuration file to make the toolset generate a figure based on that statistic.
We do not generate figures from all the statistics by default to not flood the user
with useless information.

\paragraph{Recommended workflow}

Using our toolset, the recommended workflow is the following. The author of the patch 
provides reviewers with a link to the test results. Reviewers can choose to either 
merely look at the results or repeat the whole experiment based on the configuration, 
depending on the verification effort required for the change. They can also suggest 
changes to the configuration to gain more insight into the changes.

\paragraph{Alternative use cases}

Monitoring changes in the static analysis engine is not the only use case of the 
CSA testbench scripts. We can use these scripts to tune certain parameters / heuristics of 
the analyzer to improve the results on a set of projects.

\paragraph{C-Reduce}

C-Reduce~\cite{creduce} is a tool that takes a large C, C++, or OpenCL file that has
a property of interest (such as triggering a compiler bug) and automatically produces
a much smaller C/C++ file that has the same property. We also use C-Reduce to get
minimal examples that showcase differences between two versions of the static 
analysis engine.
First, we need a file on which analysis engine versions produce different results.
This can be a different set of warnings or other statistics emitted by the engine.
These minimal examples can greatly aid our understanding of the effects of a change.
The main shortcoming of C-Reduce is the lack of support for reducing multiple 
translation units at once. We do plan to add this feature in the future.

\section{Our contributions}

Symbolic execution is a static program analysis technique that can be considered an
extension of compiler warnings. It produces diagnostics without user intervention,
based on the build system, but the underlying analysis is substantially more powerful 
than the data flow analysis present in compilers, and the reports provide the programmer
with the execution trace that leads to the bug. In turn, the time complexity of the
algorithm is exponential, thus it needs clever heuristics to be a realistic solution
for industrial projects.

- What we did:
    - Implemented CTU
    - Summary-based analysis (concept)
    - Implemented refutation
    - Implemented checkers

\section{Summary}
\label{summary}

In this paper we summarized our experiences collected while contributing to the 
state-of-the-art Clang Static Analyzer and CodeChecker products. Along with the introduction
of these tools, we also described a variety of related methods and alternative approaches. 
While we often enumerated specific examples of design decisions, we do believe
this paper will prove to be a useful resource for anyone deciding to work on static 
analysis tools, as it is easy to find analogies when analyzing other languages.

We described our tools using a holistic view on static analysis pipelines 
including build process integration, analysis framework testing,
and presentation of reports to the users. Hopefully, we demonstrated that
each stage of the pipeline is important in order to scale the analysis to large
software systems. It is important to not only think about resources like
CPU time or memory, but also developer time, which is often the most scarce resource.
We outlined some possible future work for our toolset that we consider worth pursuing.

\section*{Acknowledgement}

The project has been supported by the European Union, co-financed by the European Social Fund (EFOP-3.6.3-VEKOP-16-2017-00002).

This paper is part of dissemination of  results  of  the  Erasmus+  Key 
Action 2 (Strategic partnership for higher education)  project  No. 
2017-1-SK01-KA203-035402: "Focusing  Education on Composability, 
Comprehensibility and Correctness of Working Software". The information 
and views set out in this paper are those of the author(s) and do not 
necessarily reflect the official opinion of the European Union.  Neither 
  the  European  Union  institutions  and  bodies  nor  any  person 
acting  on their behalf may be held responsible for the use which may be 
made of the information contained therein.

\bibliographystyle{splncs04}
\bibliography{bibliography} 

\newpage
\appendix

\section{Appendix: Exploded graph}

\vfill
\begin{figure}
\centering
\includegraphics[width=\textwidth]{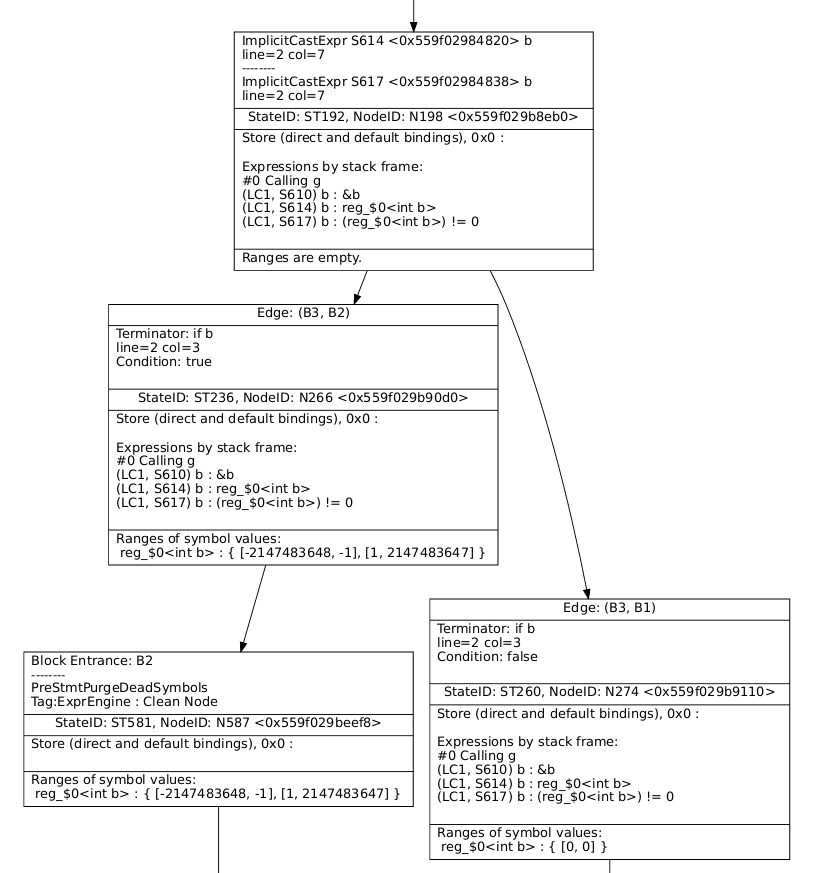}
\caption{A segment of the exploded graph generated for the sample function of Figure~\ref{fig:exploded}.}
\label{fig:realexploded}
\end{figure}
\vfill

\newpage
\section{Appendix: Cross translation unit analysis}

\vspace{100pt}
\begin{figure*}
\includegraphics[width=\textwidth]{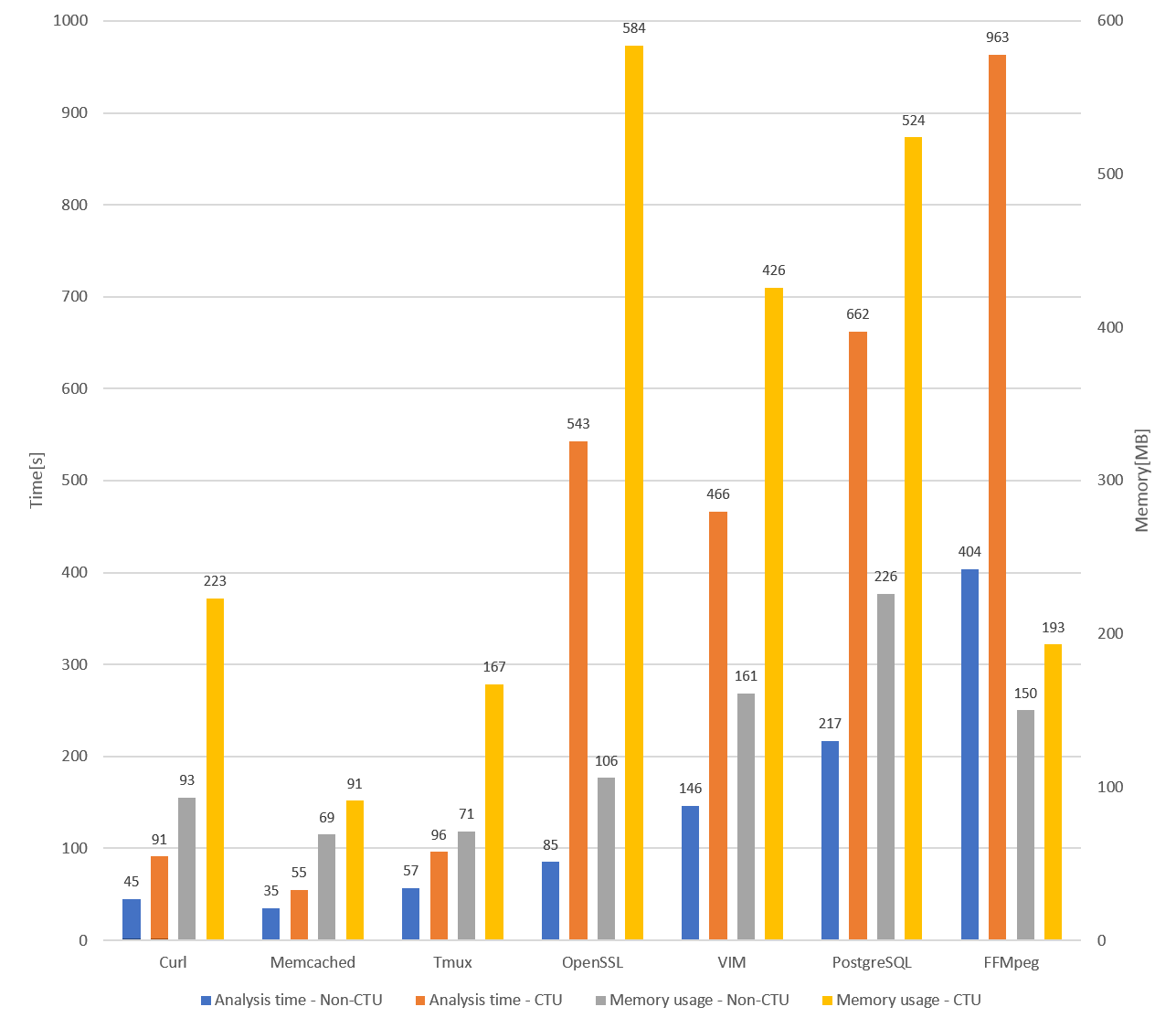}
\caption{Execution time and memory consumption characteristics of the cross-translation unit
analysis mode.}
\label{fig:ctuperf}
\end{figure*}

\begin{figure*}
\centering
\includegraphics[width=\textwidth]{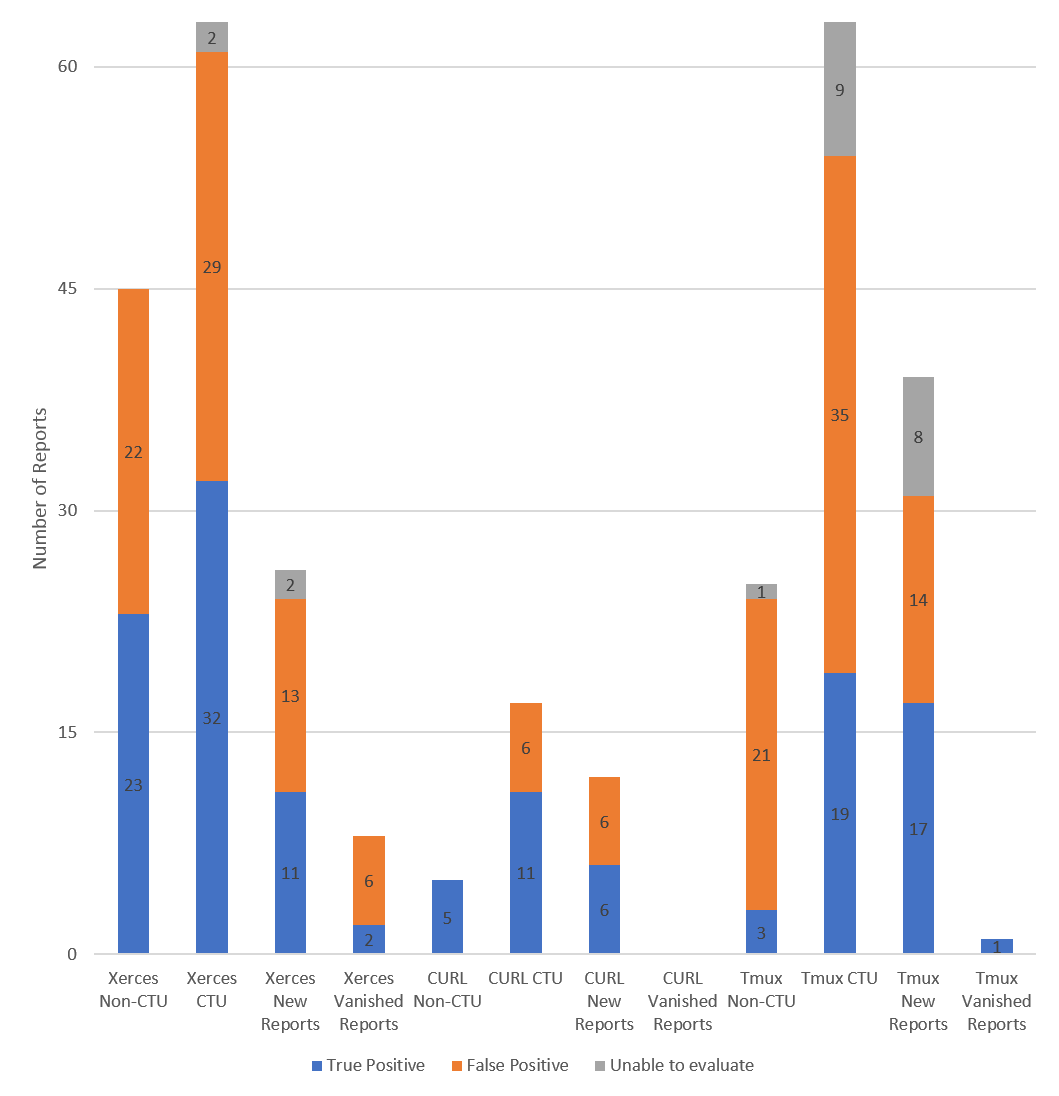}
\caption{False positive ratio using the cross-translation unit analysis mode as opposed to 
the default single translation unit mode.}
\label{fig:fpratio}
\end{figure*}

\newpage

\vspace*{50pt}
\begin{figure*}
\centering
\includegraphics[width=\textwidth]{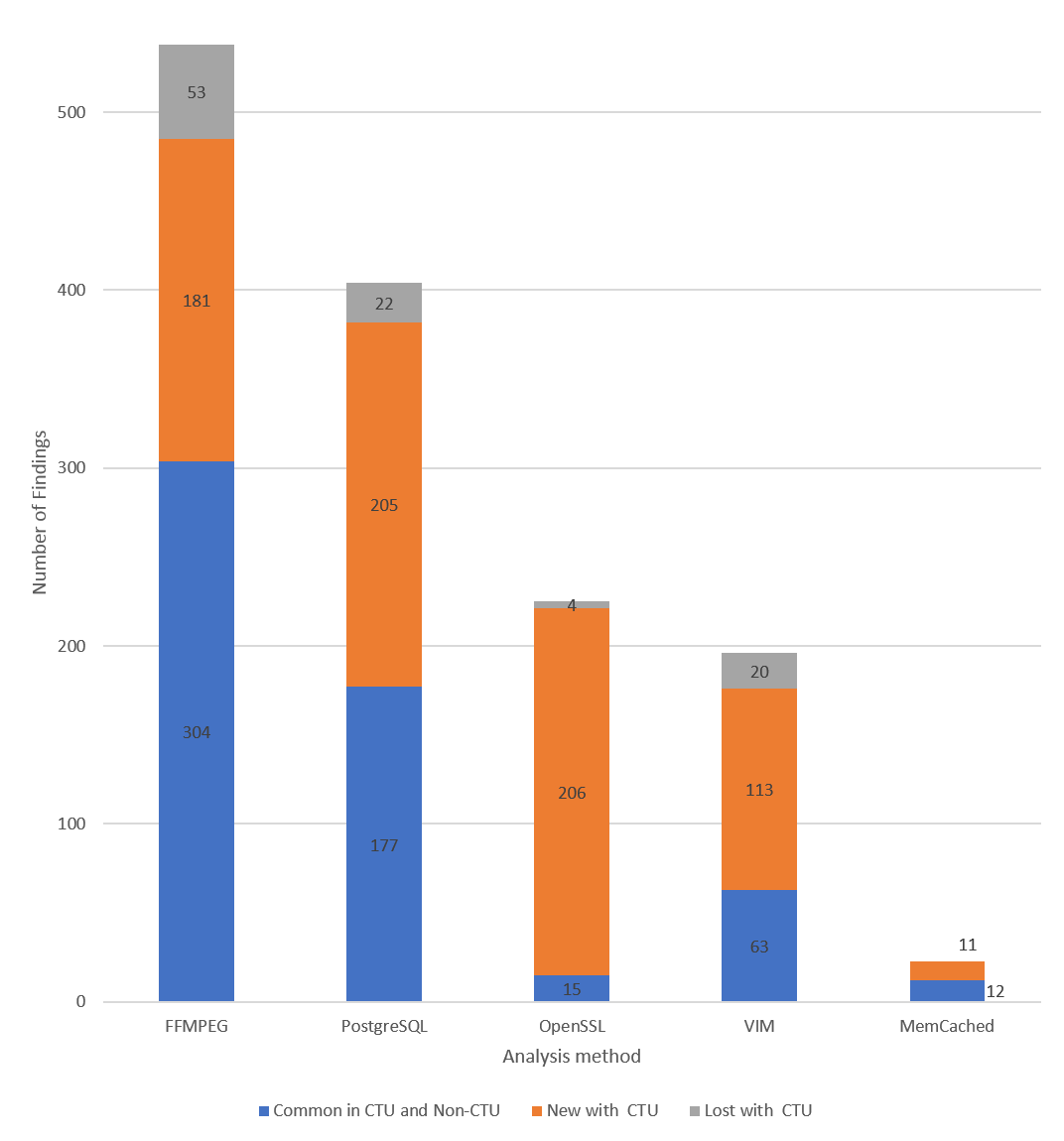}
\caption{New and lost bugs using the cross-translation unit analysis mode as opposed to the default
single translation unit mode.}
\label{fig:lostbugs}
\end{figure*}

\end{document}